\documentclass[superscriptaddress,showpacs,aps,twocolumn,prb,floatfix]{revtex4}
\usepackage[utf8]{inputenc}
\usepackage{amssymb}
\usepackage{amsmath}
\usepackage[dvips]{graphicx}
\usepackage{subfigure}
\usepackage{bm}
\usepackage{color}
\usepackage{soul}
\usepackage{gensymb}
\usepackage{wasysym}
\usepackage{xcolor}
\usepackage[export]{adjustbox}

\begin{document}
 
\title{ Polarons formation in Bi-deficient BaBiO$_3$}

\author{W. Rom\'an Acevedo}
\affiliation{Departamento de F\'{\i}sica de la Materia Condensada, GIyA-CNEA, Av. General Paz 1499, (1650) San Mart\'{\i}n, Pcia. de Buenos Aires, Argentina}
\affiliation{Instituto de Nanociencia y Nanotecnolog\'{\i}a (INN CNEA-CONICET), 1650 San Mart\'{\i}n, Argentina}

\author{S. Di Napoli}
\affiliation{Departamento de F\'{\i}sica de la Materia Condensada, GIyA-CNEA, Av. General Paz 1499, (1650) San Mart\'{\i}n, Pcia. de Buenos Aires, Argentina}
\affiliation{Instituto de Nanociencia y Nanotecnolog\'{\i}a (INN CNEA-CONICET), 1650 San Mart\'{\i}n, Argentina}

\author{F. Romano}
\affiliation{Departamento de F\'{\i}sica de la Materia Condensada, GIyA-CNEA, Av. General Paz 1499, (1650) San Mart\'{\i}n, Pcia. de Buenos Aires, Argentina}
\affiliation{Instituto de Nanociencia y Nanotecnolog\'{\i}a (INN CNEA-CONICET), 1650 San Mart\'{\i}n, Argentina}

\author{G. Rodr\'{\i}guez Ruiz}
\affiliation{Departamento de F\'{\i}sica de la Materia Condensada, GIyA-CNEA, Av. General Paz 1499, (1650) San Mart\'{\i}n, Pcia. de Buenos Aires, Argentina}
\affiliation{Instituto de Nanociencia y Nanotecnolog\'{\i}a (INN CNEA-CONICET), 1650 San Mart\'{\i}n, Argentina}

\author{P. Nukala}
\affiliation{Center for Nanoscience and Engineering, Indian Institute of Science, Bangalore, 560012, India}
\affiliation{Zernike Institute for Advanced Materials, University of Groningen, The Netherlands}

\author{C. Quinteros}
\affiliation{Zernike Institute for Advanced Materials, University of Groningen, The Netherlands}

\author{J. Lecourt}
\affiliation{CRISMAT, CNRS UMR 6508, ENSICAEN, 6 Boulevard Maréchal Juin, F-14050 Caen Cedex 4, France}

\author{U. Lüders}
\affiliation{CRISMAT, CNRS UMR 6508, ENSICAEN, 6 Boulevard Maréchal Juin, F-14050 Caen Cedex 4, France}

\author{V. Vildosola}
\affiliation{Departamento de F\'{\i}sica de la Materia Condensada, GIyA-CNEA, Av. General Paz 1499, (1650) San Mart\'{\i}n, Pcia. de Buenos Aires, Argentina}
\affiliation{Instituto de Nanociencia y Nanotecnolog\'{\i}a (INN CNEA-CONICET), 1650 San Mart\'{\i}n, Argentina}

\author{D. Rubi}
\affiliation{Departamento de F\'{\i}sica de la Materia Condensada, GIyA-CNEA, Av. General Paz 1499, (1650) San Mart\'{\i}n, Pcia. de Buenos Aires, Argentina}
\affiliation{Instituto de Nanociencia y Nanotecnolog\'{\i}a (INN CNEA-CONICET), 1650 San Mart\'{\i}n, Argentina}

\email{vildosol@tandar.cnea.gov.ar}

\begin{abstract}

BaBiO$_3$ is a charged ordered Peierls-like perovskite well known for its superconducting properties upon K or Pb doping. We present a study on the transport and electronic properties of BaBiO$_3$ perovskite with strong Bi-deficiency. We show that it is possible to synthesize BaBiO$_3$ thin layers with Bi-vacancies above 8-10\% by depositing an yttrium-stabilized zirconia capping layer. By combining transport measurements with \textit{ab initio} calculations we propose an scenario where the Bi-vacancies give rise to the formation of polarons and suggest that the electrical transport is dominated by the migration of these polarons trapped at Bi$^{3+}$ sites. Our work shows that cation vacancies engineering -hardly explored to date- appears as a promising pathway to tune the electronic and functional properties of perovskites.

\end{abstract}

\maketitle

\section{Introduction}\label{sec:intro}

Perovskite oxides display a plethora of physical properties, such as magnetism \cite{coey09}, ferroelectricity \cite{daw05} or superconductivity \cite{rao90}, which turn them very attractive for the development of devices for sensing \cite{sun20}, information storage \cite{sawa08} or computing \cite{yu_2017}, among others. The possibility of implementing these devices has boosted in the last decades as oxide thin films and heterostructures with excellent crystalline properties, chemistry and atomically controlled surfaces and interfaces have been developed. Defects are ubiquitously found in oxide thin films, oxygen vacancies being the most common ones \cite{gun20}. Cationic vacancies have been less studied due to their lower concentration in comparison with oxygen vacancies \cite{rose2021}, but they might have a significant impact on the physical properties of the films as they could strongly modify the oxide electronic structure.

BaBiO$_3$ (BBO) is a perovskite which becomes superconducting upon Pb or K doping \cite{matt98,cava98} with high critical temperatures up to $\sim$ 30 K. Pristine BBO is an insulator due to the existence of a charge ordered state comprising Bi ions with formal +3 and +5 valences \cite{cox76}. This is accompanied with structural distortions involving both the so-called breathing distortions and tiltings of BiO$_6$ octhaedra \cite{kenn06}, which reduce the crystal symmetry from a cubic to a monoclinic structure. Bi-O bond distances have been reported as 2.11 and 2.29 {\AA}, respectively \cite{kenn06}. Spectroscopic experiments have failed to resolve valence differences between Bi ions in both crystallographic sites \cite{shen89,nago92}, while theoretical work suggests a common $\approx$ +3 state, which has been rationalized in terms of spatially extended Bi-O bonds related to hybridized Bi-6s and O-2p atomic orbitals \cite{foy15,plumb16,balan17}. BBO has been proposed to behave as a topological insulator upon electron doping \cite{yan13}, and it was also theoretically suggested to present a 2D electron gas in its Bi-terminated (001) surface\cite{Vildosola2013}. Moreover, 2D superconductivity has been shown at the BBO/BaPbO$_3$ interface \cite{Meir2017,arxiv-BPO-BBO}.

Reports on BBO thin films are scarce \cite{makita97,gozar07,inumaru08,lee16,ferreyra16,ferreyra16b, zapf18,zapf19,bouwmeester19, jin2020}. Most works report on the growth and properties of BBO on SrTiO$_3$ (STO) substrates \cite{makita97,gozar07,lee16,zapf18,zapf19,bouwmeester19, jin2020}, with both materials presenting a rather large lattice mismatch of $\approx$ 12 \%. It was shown that this misfit is accommodated through the formation of an ultrathin interfacial $\delta$-Bi$_2$O$_3$ layer with a fluorite structure, which forms a coherent interface with the substrate and a semi-coherent interface with BBO \cite{zapf18,jin2020}. Recent reports indicate that the electronic structure of BBO on STO is thickness dependent and is affected by chemical and structural effects imposed by the substrate/film interface \cite{kim15,zapf19}. These results indicate that substrate effects on the BBO electronic structure might not be disregarded in thin epitaxial films.\\

With the aim of studying the effect of (B-site) Bi deficiency on the electronic and transport properties of BBO, we have developed a fabrication procedure allowing for the introduction of a large amount of Bi vacancies in stochiometric BBO thin films on silicon. From a combination of experiments and \textit{ab inito} calculations, it is proposed that Bi deficiency leads to the formation of polarons that dominate the electrical transport. Our work shows the control of cation deficiency as an emerging degree of freedom that allows tuning the BBO electronic structure and its functional properties.

\section{Experimental section}

BBO thin films were deposited by pulsed laser deposition using a Nd-YAG Spectra Physics laser with $\lambda$ = 266 nm and 850 mJ per pulse. Structural characterization of the films was performed by means of X-ray diffraction (XRD) using two Panalytical Empyrean diffractometers  with PIXcel ultrafast detectors, in standard Bragg-Brentano configuration (0.02º steps were used for 2$\theta$ scanning). Transmission electron microscopy (TEM) was performed with a Thermo Fisher Scientific Themis Z monochromated and double corrected microscope operated at 300 kV.\\
In order to get Bi-deficient BBO, we followed this procedure: first, we deposited  a stochiometric BBO film ($\approx$ 165 nm thick) on commercial Si/SiO$_2$ substrate, at a temperature and oxygen pressure of 550 ºC and 0.01 mbar, repectively. We chose this substrate in order to obtain strain-free polycrystalline films, as will be shown below. Next, we deposited (in-situ), on top of BBO, a yttrium stabilized zirconia (YSZ, $\approx$ 17 nm thick) capping layer, at room temperature, at an optimized oxygen pressure of 0.1 mbar. YSZ is a well known oxygen conductor \cite{skinner2003} but has also been reported to display high cationic conductivities \cite{kilo2000}. In our case, under the proper deposition conditions, YSZ capping favours the diffusion of Bi from BBO to YSZ, leading to an interfacial BBO layer with a strong Bi deficiency.

\begin{figure*}[t]
\centering
\includegraphics[scale=0.75]{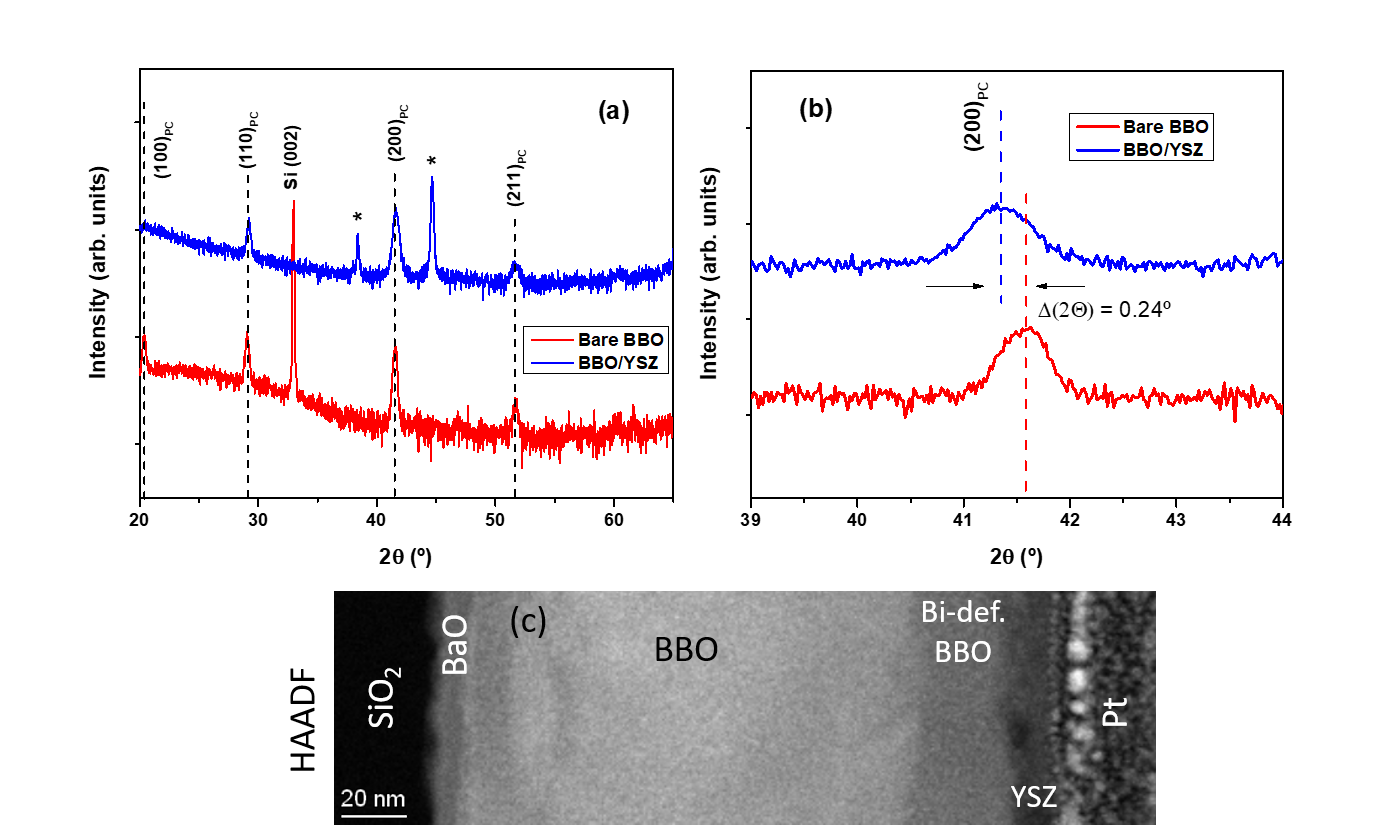}
\caption{(a) X-ray diffraction spectra corresponding to both bare and capped BBO. Peaks indexes correspond to the pseudo-cubic (PC) perovskite BBO lattice. Peaks marked with * correspond to the sample holder used for the measurement of the capped BBO sample \cite{SI}. Zero-shift corrections were made from the position of Si peaks; (b) Blow-up of the spectra displayed in (a), evidencing the shift of diffraction peaks to lower angles for capped BBO. Maxima are shifted by 0.24 º. Also, capped BBO displayed broader diffraction peaks (FWHM of $\approx$ 0.5 º and $\approx$ 0.7 º for BBO and capped BBO, respectively) with a slight asymmetry; (c) TEM HAADF image corrresponding to a BBO/YSZ bilayer on Si/SiO$_2$, with the YSZ grown at an oxygen pressure of 0.1mbar. The Pt layer on top of the YSZ layer was deposited for the lamella preparation.}
\label{Fig1}
\end{figure*}

Figure \ref{Fig1}(a) shows X-ray diffraction spectra recorded for both bare and capped BBO films. It is found in both cases the presence of peaks corresponding to the BBO perovskite structure while, for capped BBO, YSZ remains amorphous. BBO peaks were indexed considering a pseudo-cubic (PC) lattice. We extracted a cell parameter for bare BBO a$_{PC}$ = 4.339(5) {\AA}, in excellent agreement with the value corresponding to bulk BBO (4.335 {\AA}). Upon YSZ capping, there is a shift of X-ray peaks towards lower angles (see Figure \ref{Fig1}(b)), reflecting an (average) BBO unit cell parameter expansion of $\approx$ 0.5\%. This is a first indication of the formation of Bi vacancies at the BBO layer, as unit cell enlargement is a signature of B-site vacancies in perovskites \cite{keeble13,brooks15,scafetta17}. We will come back later to this issue.\\

Figure \ref{Fig1}(e) displays a HAADF-STEM cross-section of a capped BBO film. First we notice the existence of a thin BaO layer ($\approx$ 11 nm, non detectable by XRD) at the substrate/BBO interface, confirmed by the EDS maps displayed at the Suppl. Information \cite{SI}. This interfacial BaO layer has been observed before in BBO thin films \cite{ferreyra16b} and can be related to the poor sticking coefficient of Bi ions during the first growth stages \cite{li2019,zapf19}. Next, Figure \ref{Fig1}(c) displays a BBO layer divided in two zones with different contrast, the darker one ($\approx$ 30 nm thick) located at the interface with YSZ. As HAADF brightness is higher for species with high atomic number such as Bi, the darker BBO fringe is an indication of Bi-deficiency at this zone.  We notice that it is difficult to perform a precise composition quantification, as during EDS mapping we observed some grain growth together with changes in the HAADF contrast between BBO and Bi-deficient BBO \cite{SI}, indicating atomic diffusion during the measurement. With this constraint in mind, we can roughly estimate a lower bound of $\approx$ 8-10\% of Bi vacancies at the Bi-deficient BBO layer \cite{SI}. This statement is supported  from the analysis of (200)$_{PC}$ X-ray peaks performed in the Suppl. Information \cite{SI}. The (200)$_{PC}$ reflection of bare BBO can be fitted with a single Gaussian component, centered at 2$\theta$ = 41.59 º, corresponding to a bulk-like cell parameter 4.339(5) {\AA}. On the other hand, the capped BBO (200)$_{PC}$ reflection  shows a broadened peak with a slight asymmetry. The peak can be deconvoluted by assuming two Gaussian contributions centered at 2$\theta$ = 41.59 º and 2$\theta$ = 41.25 º. These two components are attributed to BBO and Bi-deficient BBO,  with cell parameters 4.339(5) {\AA} and 4.374(5) {\AA}, respectively. The latter cell parameter accounts for a $\approx$ 4 pm BBO cell expansion due to the presence of Bi-vacancies. We recall that in the case of Ti deficient SrTiO$_3$ such expansion corresponds to a 10-15\% B-cation deficiency \cite{keeble13}, allowing an additional estimation of the amount of B-site vacancies in the Bi-deficient layer of capped BBO, which is consistent with the previous one obtained from microscopy analysis.\\





Now we will focus on the electrical transport properties of our samples. BBO single crystals were reported to display an insulating, thermally activated behavior \cite{sleight15}. Transport properties in BBO thin films have been rarely published: some authors reported an "insulating behavior" \cite{inumaru08} in BBO thin films on MgO but no temperature dependent resistivity was shown, while a negative resistance coeficent, in the range 200-300 K, was displayed for BBO films on silicon \cite{ferreyra16}. We recall that the transport properties in semiconducting oxide thin films can be highly affected by the presence of defects. For instance, BaSnO$_3$ thin films were shown to present dramatic changes in their resistivity values -and the corresponding temperature behavior- depending on the films' free carriers concentration \cite{prakash16}, strongly linked to the type and number of existing defects.

\begin{figure}[h!]
\centering
\includegraphics[width=1.\columnwidth]{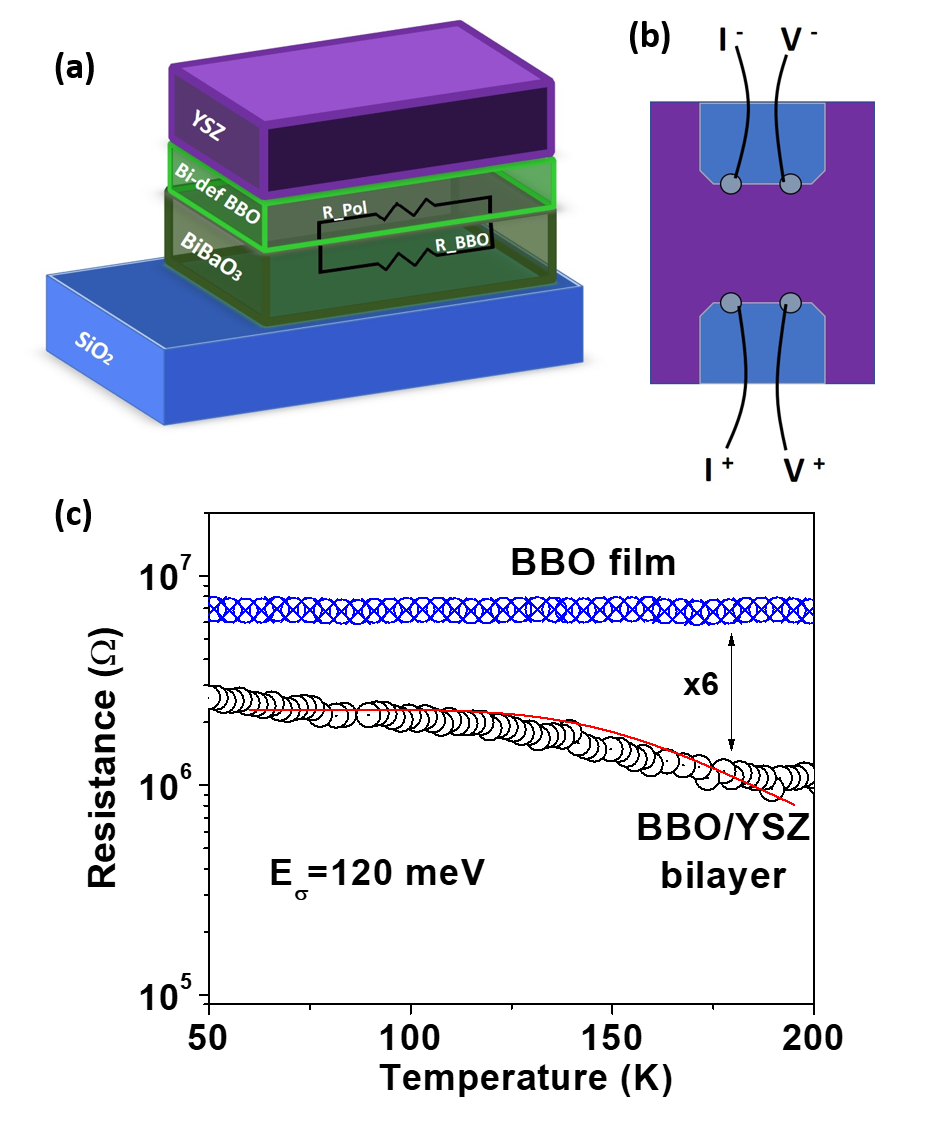}
\caption{Sketches showing (a) the Si$_2$/BBO/Bi-def. BBO/YSZ stack and (b) a top view of the contacted device. Small parts of SiO$_2$ (top layer of the substrate, blue) were left exposed in order to access the BBO and Bi-deficient BBO layers for 4-points transport measurements. Same colours are used for (a) and (b); (c) 4-points resistance as a function of the temperature recorded for a bare BBO film and a YSZ-capped heterostructure.The red line shows a fitting assuming a polaronic conduction model for the Bi-deficient BBO layer (see text for details).}
\label{Fig2}
\end{figure}

Electrical transport has been measured on the deposited structures at low temperatures (50-200 K) in a 4-point configuration using a multipurpose Quantum Design Versalab criostat. In order to gain electrical access to the capped BBO samples, we have covered some parts of the substrate by shadows masks during the deposition (Figure \ref{Fig2}(b)). Electrical contacts were fixed with silver paste. Figure \ref{Fig2}(c) displays the evolution of the 4-points resistance versus temperature for both uncapped and capped BBO films. Bare BBO films show a resistance nearly independent of the temperature -in the measured temperature range-, indicating that the expected thermally activated behavior (negative resistance coefficient) is compensated with a reduction of carriers mobility as temperature increases. We note that the mentioned behavior of the mobility with temperature implies that mobility is limited by phonons rather than by impurity scattering, this balance being very sensitive to free carriers density \cite{zhao16}. On the other hand, capped BBO shows lower resistance values -a 6-fold decrease at 175 K- together with a semiconducting-like behavior characterized by a decrease of the resistance as the temperature increases. We checked in separated control experiments that both SiO$_2$ and YSZ present resistances several orders of magnitude higher than those shown in Figure \ref{Fig2}(c) and this is also the case for the estimated resistance corresponding to the BaO layer ($>$ 10 G$\Omega$), given its reported resistivity ($\approx$ 10$^4$ $\Omega$cm at $\approx$ 700 ºC) and semiconducting-like temperature behavior \cite{dollof56}. Therefore we can safely exclude the contribution of these three layers to the electrical transport and assume that, for capped BBO, the electrical transport flows trough two parallel channels related to BBO and Bi-deficient BBO layers (see the sketch of Figure \ref{Fig2}(a)). This scenario was considered to fit the experimental data by proposing an equivalent circuit formed by a temperature independent resistance R$_{BBO}$ for the BBO layer in parallel with a polaronic transport channel, attributed to the Bi-deficient BBO layer, with a temperature dependent resistance R$_{Pol}$(T) given by \cite{jaime97}

\begin{equation}\label{eq1}
R_{Pol}(T)= \frac{K_0a}{g_d e^2 \nu_0}k_B T\exp\left(\frac{E_\sigma}{k_BT}\right) ,
\end{equation}

where k$_B$ is the Boltzmann constant, T is the absolute temperature, $e$ is the electronic charge, $a$ is the polaron hopping distance, $\nu_0$ is the jump attempt frequency -related to the phonon vibrations coupled to the polaron-, E$_\sigma$ is the activation energy for polaron hopping, g$_d$ is an adimensional factor that depends on the hopping geometry\footnote{g$_d$=$Nca^3$, where $N$ is the available sites for a jump, $c$ is the defect concentration and $a$ was defined in the text. See for instance Ref. \citenum{defects-in-solids}.} and K$_0$=L/Wt, with L, W and t being the length, width and thickness of the Bi-deficient BBO layer, respectively. 

The R(T) for the bilayer is thus given by

\begin{equation}\label{eq2}
R(T)= \frac{R_{BBO} R_{Pol}(T)}{R_{BBO} + R_{Pol}(T)}.
\end{equation}

The fitting of R(T) is shown with a red full line in Figure \ref{Fig2}(c) and displays a reasonable agreement with the experimental data for E$_\sigma$=0.12 eV and $K_0\frac{ak_B} {g_de^2\nu_0} =5.0  \Omega/K$. 

If we take $\nu_0$=1.710$^{13}$ Hz (the breathing mode vibration frequency\cite{PhysRevLett.55.426}) , a=0.437nm -corresponding to this perovskite cell parameter-, g$_d$=3/8 \footnote{g$_d$=3/8 taking N equal to 6 and $c$ is estimated considering 1 polaron in the supercell of volume $(4a)^3/4$.} and $K_0$=0.067 nm$^{-1}$ (L/W$\approx$2 given by our sample geometry and t=30 nm, as extracted from the TEM analysis of Figure \ref{Fig1}) we obtain a theoretical value $K_0\frac{ak_B} {g_de^2\nu_0} =2.5 \Omega/K$, in very good agreement with the fitted value, giving consistency to our analysis.

\begin{figure*}[ht!]
\includegraphics[width=1.85\columnwidth]{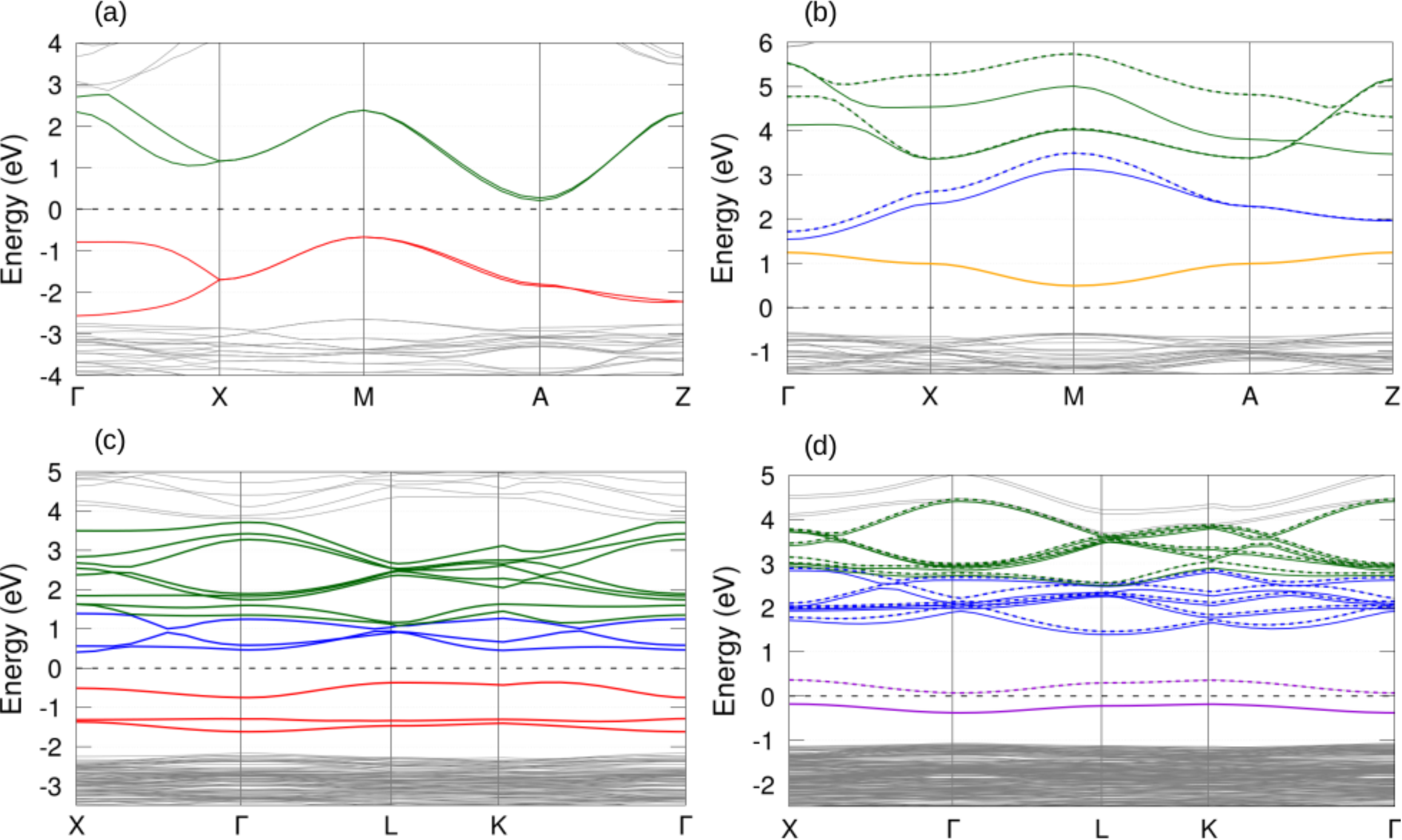}
\caption{Spin-polarized band structure of (a) perfect BaBiO$_3$ and Bi-deficient BaBiO$_3$  with (b) 25\,\%, (c) 12.5\,\% and (d) 18.75\% of Bi-vacancies. Solid lines correspond to spin-up projection and dashed lines to spin-down. In (a) and (c) spin-up and spin-down bands are degenerate. Red bands correspond to Bi$^{+3}$, dark-green ones to original Bi$^{+5}$ atoms, and blue to created bipolarons. 
The orange band in (b) represents the hole trapped at the oxygen cloud centered at the Bi-vacancy sketched in Figure \ref{1vac-charge}~(a) and
the violet bands in (d) correspond to the charge/hole polaron originated around Bi$_{12}$ (see text and Figure~\ref{3vac-PDOS} (b))}   
\label{bands}
\end{figure*}

We recall that it has been theoretically proposed that, upon hole doping, BBO remains insulating due to the coupling between holes and phonons \cite{franchini09} and the subsequent formation of polarons. This is likely the case of Bi-deficient BBO, where the holes for polaron formation are provided by Bi vacancies, as it will be analyzed in the next section by means of first principles calculations.

\section{DFT calculations}

To study the different physical mechanisms that could trigger the transport properties of Bi-deficient BaBiO$_3$, we perform \textit{ab initio} calculations within  the framework of Density Functional Theory (DFT) and the projector  augmented wave (PAW) method~\cite{PAW}, as implemented in the Vienna \textit{ab initio} package (VASP)~\cite{VASP,PAW-VASP}. We explicitly treat 10 valence electrons for Ba (5s$^2$5p$^6$6s$^2$), 15 for Bi (5d$^{10}$6s$^2$6p$^3$)  and 6 for O (2s$^2$2p$^4$). 
To properly describe the electronic structure of charge ordered and semiconducting BaBiO$_3$ -specially in the presence of localized defects-, it is necessary to consider long range exchange interactions that are poorly described by the standard local and semilocal functionals as the local density approximation (LDA) or the general gradient approximation (GGA).  Therefore, we use the semilocal functionals developed by Perdew, Burke and Ernzerhof (PBE)~\cite{PBE96}  and we add the hybrid functional including a fraction of non-local Hartree-Fock exchange interactions through the  Heyd-Scuseria-Ernzerhof (HSE06) functionals~\cite{2003-HSE1,2006-HSE2} in all the calculations. \\
The charge-ordered phase of BaBiO$_3$ is represented formally by the  Ba$_2^{+2}$Bi$^{+3}$Bi$^{+5}$O$_6^{-2}$ formula unit. The two Bi species are ordered in an alternated way within a monoclinic structure~\cite{Cox1976969}.  To compute the electronic properties of perfect BaBiO$_3$ and  Bi-deficient BaBi$_{0.75}$O$_3$, containing a 25$\%$ of Bi-vacancies, we use a supercell rotated 45$^{\circ}$ with respect to the simple cubic single perovskite and which contains 4 Bi sites in total, 2 per layer and two planes in the $z$-direction. To explore lower concentrations of Bi-vacancies we also perform calculations in a 2x2x2 FCC supercell containing 80 atoms~\footnote{To build this FCC supercell we begin with a 2x2x2 supercell of the simple cubic perovskite containing 40 atoms in total and we perform several molecular dynamics steps, followed by relaxation, to stabilize the breathing distortion driven by the charge disproportionation. Once this distortion agree well with the experimental one, we build up again a 2x2x2 supercell, described within the FCC Bravais lattice.}, as the one used in Ref.~\onlinecite{franchini09}. This larger supercell contains 16 Bi-atoms per unit cell and we account for defect concentrations of 12.5$\%$ (2 Bi-vacancies) and 18.75$\%$ (3 Bi-vacancies).\\
The HSE calculations within the smaller unit cell are performed using a 450~eV energy cutoff in the plane waves basis and to evaluate the integrals within the Brillouin Zone (BZ), a 4$\times$4$\times$3 Monkhorst-Pack $k$-point grid is employed. The corresponding values for the energy cutoff and $k$-points grid in the larger FCC supercell are 400~eV and 2$\times$2$\times$2, respectively. The experimental lattice parameters~\cite{Cox1976969} are used in all the calculations and internal structural relaxations are performed until the forces on each ion are less than 0.02~eV/\AA. \\

For perfect BaBiO$_3$, our calculations predict the semiconducting state with a band gap of 0.90 eV, exhibiting the charge disproportionation between the two different species, Bi$^{+3}$ and Bi$^{+5}$ in agreement with previous calculations \cite{PhysRevB.81.085213}. In Figure \ref{bands}(a) we show the corresponding bandstructure where the full 6s bands of Bi$^{+3}$ are depicted in red and the empty ones of Bi$^{+5}$ are in dark green. It is important to mention that the Bi-O bonds can be described by spatially extended hybridized 6s-2p orbitals and all the bands present a strong s-p hybridization. The color coding used in Figure \ref{bands} and in the following ones is just a guide to the eye.\\


Distinct kinds of defects appear in the presence of Bi vacancies. It is worth to mention that several molecular dynamics steps, followed by relaxation, were necessary to stabilize these defects in order to avoid spurious metallic solutions~\cite{franchini09}. In the following, we describe the results obtained for the different Bi vacancies concentrations considered. 

\begin{figure}[ht!]
\includegraphics[width=.85\columnwidth]{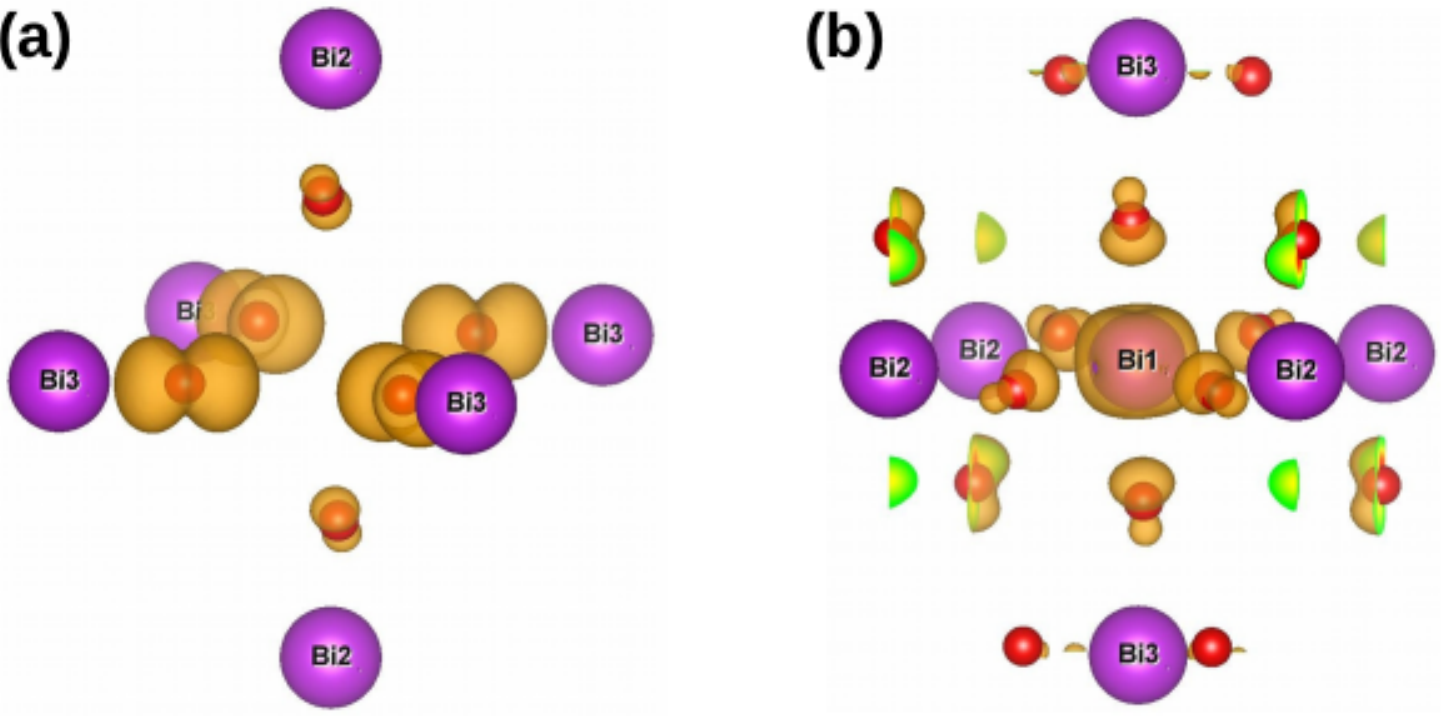}
\caption{Charge density plots of Bi-deficient BaBiO$_3$ with a 25\% of Bi-vacancies. Panel (a) shows the hole trapped at the oxygen cloud around the vacancy, correspoding to the orange band depicted in Figure~ \ref{bands} b). The charge density shown in (b) corresponds to the bipolaron at the remaining Bi$^{+3}$ site (blue band in Figure~\ref{bands}~b).}
\label{1vac-charge}
\end{figure}

The system with 25 $\%$ of Bi vacancy concentration has been simulated in the small cell where one out of four Bi is withdrawn. Independently of the site of the vacancy, whether it is at a  Bi$^{+5}$ or Bi$^{+3}$ site, the resulting electronic structure is very similar. For instance, if the vacancy is at a Bi$^{+5}$ site, in order to compensate the lacking five electrons, each of the remaining two Bi$^{+3}$ liberates two electrons, so that the originally full 6s bands become empty. These two trapped holes at each $Bi^{+3}$ form a bipolaron. Since the remaining Bi$^{+5}$ site has its 6s band already empty, the other missing electron turns into a hole trapped at the surrounding oxygen cloud next to the Bi vacancy. On the other way around, if the vacancy is at a Bi$^{+3}$ site, the lacking three electrons give rise to a bipolaron at the remaining Bi$^{+3}$ site and a hole trapped at the oxygen p-bands surrounding the vacancy. In Figure \ref{bands} (b) we show the electronic bandstructure, with the orange narrow band corresponding to holes trapped at the oxygen cloud around the vacancy, whose charge density is depicted in Figure \ref{1vac-charge}(a). The bands in blue are the bipolarons with the corresponding charge density shown in Figure \ref{1vac-charge}(b).

As mentioned before, the system with 12.5 $\%$ and 18.5~$\%$ Bi vacancy concentrations has been simulated with a supercell containing 16 Bi sites. In the case with an even amount of vacancies, the kind of electronic defects are only of the bipolaron type.  \\

In Figure \ref{bands} (c), we show the calculated electronic bandstructure obtained for the 12.5~$\%$ concentration, where there are two vacancies at Bi$^{+3}$ sites in the unit cell. The three occupied bands in red are the full 6s bands of the remaining Bi$^{+3}$, the three bands in blue are those of the bipolarons, and, at higher energies, the eight 6s bands of the Bi$^{+5}$ are in dark-green. In the case of even number of vacancies, the system is non-spin polarized so that the bands are doubly degenerate.

When the Bi vacancy concentration is 18.75~$\%$, there are three vacancies in the supercell. For example, if two of the vacancies are at a Bi$^{+3}$ site and the other one at a  Bi$^{+5}$, there are eleven electrons missing. The six remaining Bi$^{+3}$ sites can provide this charge. That is, five of these Bi$^{+3}$ sites turn into Bi$^{+5}$ by trapping two holes each giving rise to five bipolarons (blue bands in Figure \ref{bands} (d)). The remaining Bi$^{+3}$ keeps one electron forming a polaron with a localized spin-polarized charge (violet band in Figure \ref{bands} (d)).\\ 


In order to analyse the role played by each type of defects into the transport properties we proceed to estimate the activated diffusion barrier E$_{\sigma}$ in each case.  


\begin{figure}[ht!]
\includegraphics[width=.85\columnwidth]{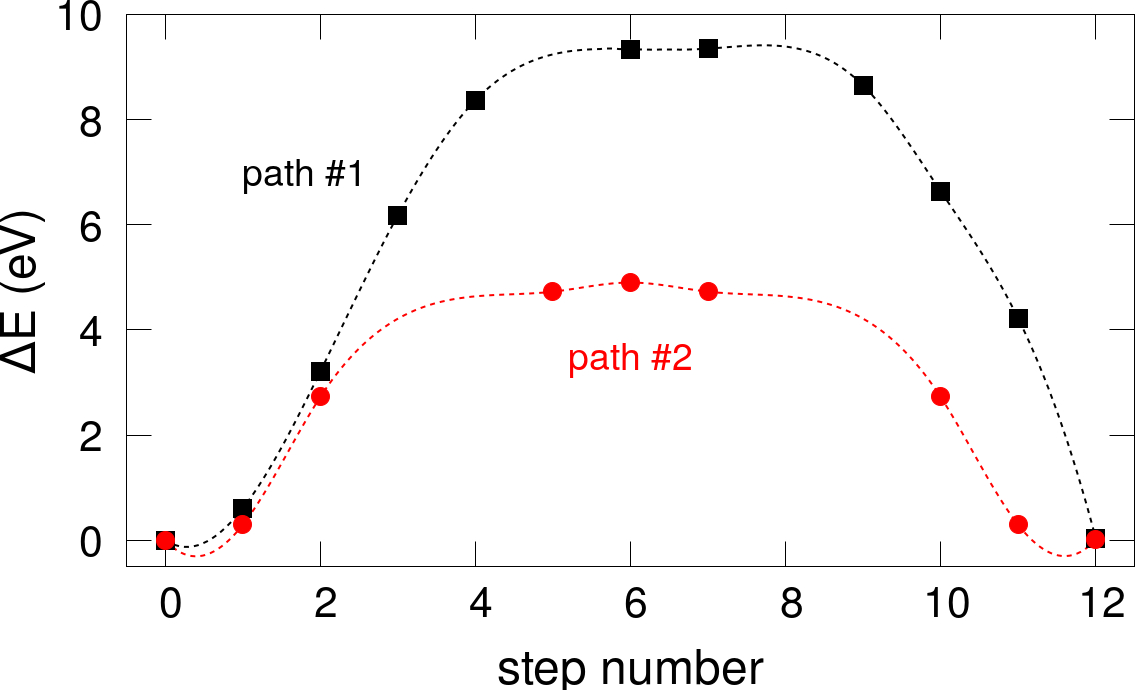}
\caption{Activation barrier of the hole polaron trapped at the oxygen $p$-bands that surround the Bi vacancy in the highly doped case, i.e. the 25$\%$ concentration of Bi-vacancies. The barrier is calculated using the Nudged elastic bands (NEB,) taking into account two different migration pathways (see text for details).}
\label{1vac-barrier}
\end{figure}

The hole polaron trapped at the $p$-bands of the oxygens that surround the Bi vacancy in the highly doped case (25$\%$ vacancy concentration), is a mixed type of defect where the Bi vacancy is bound to a hole electronic cloud. Nudged elastic bands (NEB)\cite{NEB1} method has been used to estimate the activation barrier, taking into account two different migration pathways. The first one considers the migration of the defect within the $ab$-plane, that is in the plane that contains the part of the hole shared by the 4 in-plane oxygen atoms. The second considered path was with the defect travelling along the (111)-($\bar1 \bar1 \bar1$) direction, with the inflection point located at an optimized height. That is, the defect travelling from the (1/2,1/2,1/2) to the (0,0,0) position, through (1/4,1/4,h$_0$), being h$_0$ the optimized distance in the $z$-direction from the $ab$-plane. We obtain E$_{\sigma}$=9.35~eV and 4.9~eV, respectively. In view of these results, we conclude that the contribution to the electronic  conductivity of these type of mixed defects is negligible.\\

\begin{figure}[ht!]
\includegraphics[width=.85\columnwidth]{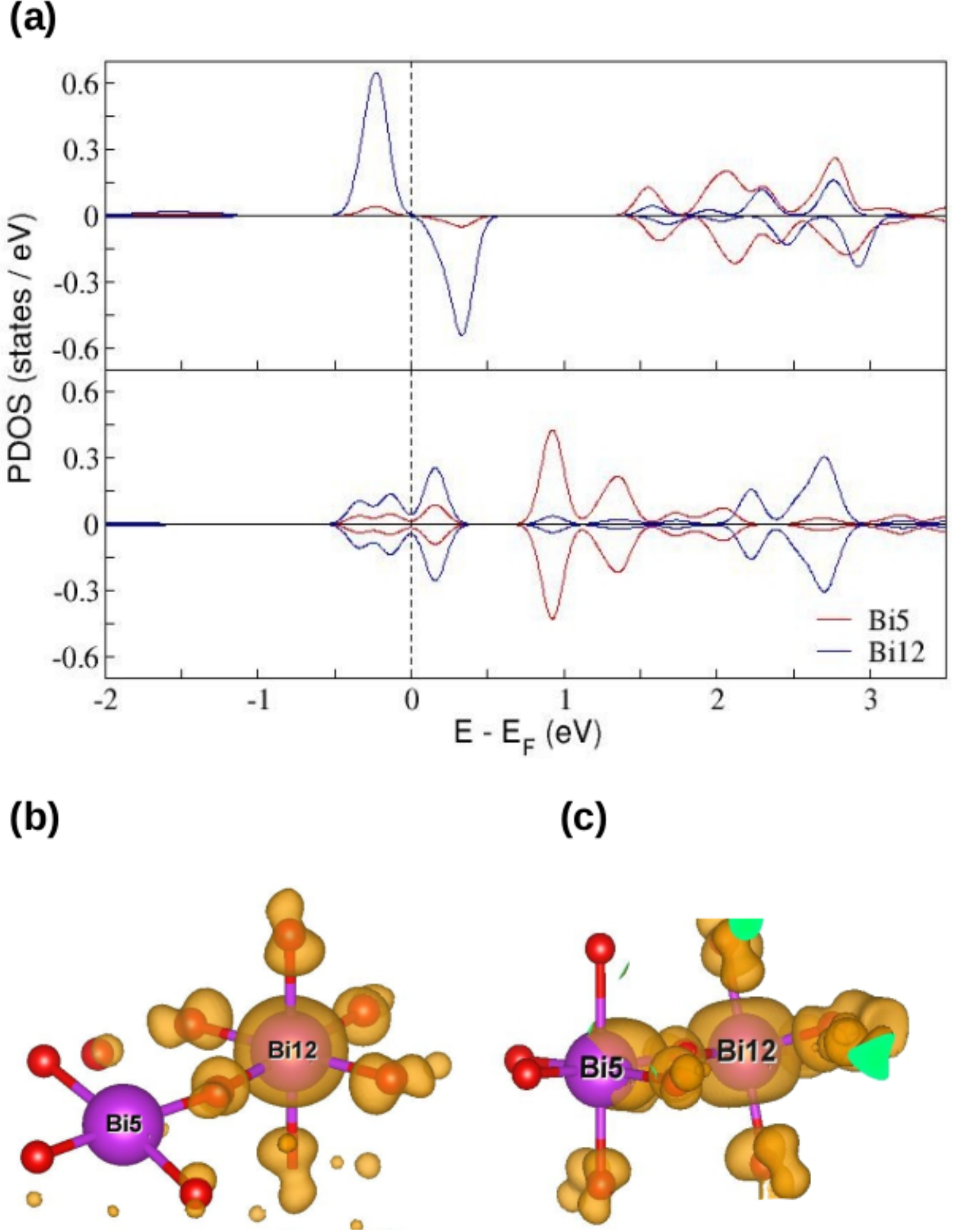}
\caption{(a-top panel) PDOS projected onto the $s$-orbitals of two selected Bi atoms, the one with the polaron, i.e. Bi$_{12}$ and one of its nearest Bi neighbors, i.e. Bi$_5$; (a-bottom) PDOS projected onto the same Bi atoms for the intermediate situation where the polaronic charge is half-transfered between neighboring sites; (b) Charge-density plot corresponding to the localized state close to E$_F$ shown in the top panel of (a), i.e. the polaron centered at Bi$_{12}$; (c) Idem (b) after the static perturbation where the polaronic charge is shared with Bi$_5$.}   
\label{3vac-PDOS}
\end{figure}

To estimate the maximun of the energy barrier of the polaron and bipolaron trapped at Bi$^{+3}$ sites (violet  and blue bands of Figure \ref{bands}, respectively), we perform a static calculation where the lattice distortion coupled to the defect is transferred to an intermediate situation that forces the trapped charge to be shared between neighboring Bi sites. \\
For the case with 18.75~$\%$ Bi vacancy concentration, in the top panel of Figure \ref{3vac-PDOS} (a), we show the partial density of states (PDOS) projected onto the s-orbitals of two Bi neighboring sites, one of them trapping the polaron (labeled as Bi$_{12}$) and the other one belongs to a  Bi$^{+5}$, named Bi$_{5}$. The corresponding charge density associated with the polaron is shown in Figure \ref{3vac-PDOS} (b). In the bottom panel of Figure \ref{3vac-PDOS} (a), we plot the PDOS projected on the same Bi sites but in the situation where the polaron charge is shared in equal parts between them. In Figure \ref{3vac-PDOS} (c) we show the related charge density. The energy difference between the initial and the intermediate situation  provides an estimation of E$_{\sigma}$, that is 0.54~eV. When this approach is done to estimate E$_{\sigma}$ for the bipolaron migration, where now two holes are hopping simultaneously, we obtain E$_{\sigma}$=0.60~eV.\\  

Interestingly, it is not surprising that this energy scale is related to the electron-phonon coupling in BBO systems, specially with the breathing phonon modes that are strongly correlated to Bi$^{+5}$- Bi$^{+3}$ charge ordering. The value around 0.5 eV  for E$_{\sigma}$ is of the same order of magnitude as the energy shift induced in the electronic bands when frozen phonon pertubations for the breathing mode are performed for bulk BBO\cite{Kotliar2013}. However, as it has been recently reported in Ref. \citenum{arxiv-BPO-BBO}, this value decreases to $\sim$ 0.2 eV in confined situations as the one described for the BBO/BaPbO$_3$ interface. So, it is likely that the value of $\sim$ 0.5 eV obtained for the bulk supercell described here is overestimated since it does not take into account the confinement effects present in the real samples. Therefore, we consider that the actual activation energy might be close to the 0.12 eV obtained from the transport experiments and that the migration of polarons trapped at Bi$^{+3}$ sites plays an important role in the increment of the electrical conductivity with the temperature, experimentally observed in our Bi deficient films.

The existence of polarons or bipolarons trapped at Bi sites in BBO was already theoretically described in Ref.~\citenum{franchini09}. The difference is that in that work the doping is caused by K substitution at Ba sites.  They show that local lattice relaxations are sufficiently large to screen the localized holes and that hole doped BBO remains semiconducting upon moderate hole doping. In our work, we show that the level of doping with Bi vacancies is 3 to 5 times larger than with K. Notwithstanding this, it is remarkable that the system remains semiconducting for all the concentrations studied.\\    

\section{Conclusions}

We have shown that it is possible to synthesize thin layers of BBO perovskite with a large amount of Bi-vacancies. Transport properties were modelled with a polaronic model and the existence of these defects was theoretically validated. Our DFT calculations performed for Bi-deficient BBO supercells show that Bi vacancies give rise to different type of polarons, depending on the vacancies concentration. After the Bi vacancy is formed, the missing electrons induce polaron and/or bipolarons that consist in one or two holes trapped at some of the remaining Bi sites, or even at the oxygen octahedron surrounding the vacancy for large vacancies concentrations.\\ We conclude that hole doped BBO remains semiconducting despite the large hole concentration induced by the Bi deficiency. We finally estimate the activated diffusion barrier for all these type of defects to analyse their contribution to the transport properties. We show that the calculated energy barriers of the charges trapped at Bi sites are correlated to the electron-phonon coupling energy scale. The calculated activated barriers  for the simulated bulk supercells are somewhat larger than the one obtained from the fit of the transport experiments. In view of previous theoretical results for other BBO interfaces, these larger values are expected to decrease in confined situations -such as the Bi-deficient BBO layer studied in this work- and match the experimentally extracted activation energies. We finally conclude that the polarons trapped at Bi sites strongly contribute to the electrical conductivity of Bi-deficient BBO and, in a more general framework, that the control of Bi vacancies can be used as a tool for tuning the electronic properties of this perovskite.  \\

\section{Acknowledgments}
The authors received financial support from PICTs 2015-0869, 2016-0867,  2017-1836 and 2019-02128 of ANPCyT, Argentina. Also, funding from EU-H2020-RISE project "MELON" (SEP-2106565560) is acknowledged. DR thanks the Nederlandse Organisatie voor Wetenschappelijk Onderzoek for a visiting grant, project 040.11.735, "Memcapacitive elements for cognitive devices". We thank D. Vega, from CAC-CNEA, for some of the XRD measurements.

\bibliography{biblio_gases2D-oxidos}

\begin{thebibliography}{57}
\expandafter\ifx\csname natexlab\endcsname\relax\def\natexlab#1{#1}\fi
\expandafter\ifx\csname bibnamefont\endcsname\relax
  \def\bibnamefont#1{#1}\fi
\expandafter\ifx\csname bibfnamefont\endcsname\relax
  \def\bibfnamefont#1{#1}\fi
\expandafter\ifx\csname citenamefont\endcsname\relax
  \def\citenamefont#1{#1}\fi
\expandafter\ifx\csname url\endcsname\relax
  \def\url#1{\texttt{#1}}\fi
\expandafter\ifx\csname urlprefix\endcsname\relax\def\urlprefix{URL }\fi
\providecommand{\bibinfo}[2]{#2}
\providecommand{\eprint}[2][]{\url{#2}}

\bibitem[{\citenamefont{Coey et~al.}(2009)\citenamefont{Coey, Viret, and von
  Molnár}}]{coey09}
\bibinfo{author}{\bibfnamefont{J.}~\bibnamefont{Coey}},
  \bibinfo{author}{\bibfnamefont{M.}~\bibnamefont{Viret}}, \bibnamefont{and}
  \bibinfo{author}{\bibfnamefont{S.}~\bibnamefont{von Molnár}},
  \bibinfo{journal}{Advances in Physics} \textbf{\bibinfo{volume}{58}},
  \bibinfo{pages}{571} (\bibinfo{year}{2009}).

\bibitem[{\citenamefont{Dawber et~al.}(2005)\citenamefont{Dawber, Rabe, and
  Scott}}]{daw05}
\bibinfo{author}{\bibfnamefont{M.}~\bibnamefont{Dawber}},
  \bibinfo{author}{\bibfnamefont{K.~M.} \bibnamefont{Rabe}}, \bibnamefont{and}
  \bibinfo{author}{\bibfnamefont{J.~F.} \bibnamefont{Scott}},
  \bibinfo{journal}{Rev. Mod. Phys.} \textbf{\bibinfo{volume}{77}},
  \bibinfo{pages}{1083} (\bibinfo{year}{2005}).

\bibitem[{\citenamefont{Rao}(1990)}]{rao90}
\bibinfo{author}{\bibfnamefont{C.~N.~R.} \bibnamefont{Rao}},
  \bibinfo{journal}{Ferroelectrics} \textbf{\bibinfo{volume}{102}},
  \bibinfo{pages}{297} (\bibinfo{year}{1990}).

\bibitem[{\citenamefont{Shellaiah and Sun}(2020)}]{sun20}
\bibinfo{author}{\bibfnamefont{M.}~\bibnamefont{Shellaiah}} \bibnamefont{and}
  \bibinfo{author}{\bibfnamefont{K.~W.} \bibnamefont{Sun}},
  \bibinfo{journal}{Chemosensors} \textbf{\bibinfo{volume}{8}},
  \bibinfo{pages}{55} (\bibinfo{year}{2020}).

\bibitem[{\citenamefont{Sawa}(2008)}]{sawa08}
\bibinfo{author}{\bibfnamefont{A.}~\bibnamefont{Sawa}},
  \bibinfo{journal}{Materials Today} \textbf{\bibinfo{volume}{11}},
  \bibinfo{pages}{28} (\bibinfo{year}{2008}).

\bibitem[{\citenamefont{Yu}(2017)}]{yu_2017}
\bibinfo{author}{\bibfnamefont{S.}~\bibnamefont{Yu}},
  \emph{\bibinfo{title}{Neuro-inspiring computing using resistive synaptic
  devices}} (\bibinfo{publisher}{Springer International Publishing},
  \bibinfo{year}{2017}).

\bibitem[{\citenamefont{Gunkel et~al.}(2020)\citenamefont{Gunkel, Christensen,
  Chen, and Pryds}}]{gun20}
\bibinfo{author}{\bibfnamefont{F.}~\bibnamefont{Gunkel}},
  \bibinfo{author}{\bibfnamefont{D.~V.} \bibnamefont{Christensen}},
  \bibinfo{author}{\bibfnamefont{Y.~Z.} \bibnamefont{Chen}}, \bibnamefont{and}
  \bibinfo{author}{\bibfnamefont{N.}~\bibnamefont{Pryds}},
  \bibinfo{journal}{Appl. Phys. Lett.} \textbf{\bibinfo{volume}{116}},
  \bibinfo{pages}{120505} (\bibinfo{year}{2020}).

\bibitem[{\citenamefont{Rose et~al.}(2021)\citenamefont{Rose, Šmíd, Vorokhta,
  Slipukhina, Andrä, Bluhm, Duchoň, Ležaić, Chambers, Dittmann
  et~al.}}]{rose2021}
\bibinfo{author}{\bibfnamefont{M.-A.} \bibnamefont{Rose}},
  \bibinfo{author}{\bibfnamefont{B.}~\bibnamefont{Šmíd}},
  \bibinfo{author}{\bibfnamefont{M.}~\bibnamefont{Vorokhta}},
  \bibinfo{author}{\bibfnamefont{I.}~\bibnamefont{Slipukhina}},
  \bibinfo{author}{\bibfnamefont{M.}~\bibnamefont{Andrä}},
  \bibinfo{author}{\bibfnamefont{H.}~\bibnamefont{Bluhm}},
  \bibinfo{author}{\bibfnamefont{T.}~\bibnamefont{Duchoň}},
  \bibinfo{author}{\bibfnamefont{M.}~\bibnamefont{Ležaić}},
  \bibinfo{author}{\bibfnamefont{S.~A.} \bibnamefont{Chambers}},
  \bibinfo{author}{\bibfnamefont{R.}~\bibnamefont{Dittmann}},
  \bibnamefont{et~al.}, \bibinfo{journal}{Advanced Materials}
  \textbf{\bibinfo{volume}{33}}, \bibinfo{pages}{2004132}
  (\bibinfo{year}{2021}).

\bibitem[{\citenamefont{Mattheiss et~al.}(1988)\citenamefont{Mattheiss, Gyorgy,
  and Johnson}}]{matt98}
\bibinfo{author}{\bibfnamefont{L.~F.} \bibnamefont{Mattheiss}},
  \bibinfo{author}{\bibfnamefont{E.~M.} \bibnamefont{Gyorgy}},
  \bibnamefont{and} \bibinfo{author}{\bibfnamefont{D.~W.}
  \bibnamefont{Johnson}}, \bibinfo{journal}{Phys. Rev. B}
  \textbf{\bibinfo{volume}{37}}, \bibinfo{pages}{3745} (\bibinfo{year}{1988}).

\bibitem[{\citenamefont{Cava et~al.}(1988)\citenamefont{Cava, Batlogg,
  Krajewski, Farrow, Rupp~Jr, White, Short, Peck, and Kometani}}]{cava98}
\bibinfo{author}{\bibfnamefont{R.~J.} \bibnamefont{Cava}},
  \bibinfo{author}{\bibfnamefont{B.}~\bibnamefont{Batlogg}},
  \bibinfo{author}{\bibfnamefont{J.~J.} \bibnamefont{Krajewski}},
  \bibinfo{author}{\bibfnamefont{R.}~\bibnamefont{Farrow}},
  \bibinfo{author}{\bibfnamefont{L.~W.} \bibnamefont{Rupp~Jr}},
  \bibinfo{author}{\bibfnamefont{A.~E.} \bibnamefont{White}},
  \bibinfo{author}{\bibfnamefont{K.}~\bibnamefont{Short}},
  \bibinfo{author}{\bibfnamefont{J.~F.} \bibnamefont{Peck}}, \bibnamefont{and}
  \bibinfo{author}{\bibfnamefont{T.}~\bibnamefont{Kometani}},
  \bibinfo{journal}{Nature} \textbf{\bibinfo{volume}{332}},
  \bibinfo{pages}{814} (\bibinfo{year}{1988}).

\bibitem[{\citenamefont{Cox and Sleight}(1976{\natexlab{a}})}]{cox76}
\bibinfo{author}{\bibfnamefont{D.}~\bibnamefont{Cox}} \bibnamefont{and}
  \bibinfo{author}{\bibfnamefont{A.}~\bibnamefont{Sleight}},
  \bibinfo{journal}{Solid State Communications} \textbf{\bibinfo{volume}{19}},
  \bibinfo{pages}{969} (\bibinfo{year}{1976}{\natexlab{a}}).

\bibitem[{\citenamefont{Kennedy et~al.}(2006)\citenamefont{Kennedy, Howard,
  Knight, Zhang, and Zhou}}]{kenn06}
\bibinfo{author}{\bibfnamefont{B.~J.} \bibnamefont{Kennedy}},
  \bibinfo{author}{\bibfnamefont{C.~J.} \bibnamefont{Howard}},
  \bibinfo{author}{\bibfnamefont{K.~S.} \bibnamefont{Knight}},
  \bibinfo{author}{\bibfnamefont{Z.}~\bibnamefont{Zhang}}, \bibnamefont{and}
  \bibinfo{author}{\bibfnamefont{Q.}~\bibnamefont{Zhou}},
  \bibinfo{journal}{Acta Crystallographica Section B}
  \textbf{\bibinfo{volume}{62}}, \bibinfo{pages}{537} (\bibinfo{year}{2006}).

\bibitem[{\citenamefont{Shen et~al.}(1989)\citenamefont{Shen, Lindberg, Wells,
  Dessau, Borg, Lindau, Spicer, Ellis, Kwei, Ott et~al.}}]{shen89}
\bibinfo{author}{\bibfnamefont{Z.-X.} \bibnamefont{Shen}},
  \bibinfo{author}{\bibfnamefont{P.~A.~P.} \bibnamefont{Lindberg}},
  \bibinfo{author}{\bibfnamefont{B.~O.} \bibnamefont{Wells}},
  \bibinfo{author}{\bibfnamefont{D.~S.} \bibnamefont{Dessau}},
  \bibinfo{author}{\bibfnamefont{A.}~\bibnamefont{Borg}},
  \bibinfo{author}{\bibfnamefont{I.}~\bibnamefont{Lindau}},
  \bibinfo{author}{\bibfnamefont{W.~E.} \bibnamefont{Spicer}},
  \bibinfo{author}{\bibfnamefont{W.~P.} \bibnamefont{Ellis}},
  \bibinfo{author}{\bibfnamefont{G.~H.} \bibnamefont{Kwei}},
  \bibinfo{author}{\bibfnamefont{K.~C.} \bibnamefont{Ott}},
  \bibnamefont{et~al.}, \bibinfo{journal}{Phys. Rev. B}
  \textbf{\bibinfo{volume}{40}}, \bibinfo{pages}{6912} (\bibinfo{year}{1989}).

\bibitem[{\citenamefont{Nagoshi et~al.}(1992)\citenamefont{Nagoshi, Suzuki,
  Fukuda, Ueki, Tokiwa, Kiruchi, Syono, and Tachiki}}]{nago92}
\bibinfo{author}{\bibfnamefont{M.}~\bibnamefont{Nagoshi}},
  \bibinfo{author}{\bibfnamefont{T.}~\bibnamefont{Suzuki}},
  \bibinfo{author}{\bibfnamefont{Y.}~\bibnamefont{Fukuda}},
  \bibinfo{author}{\bibfnamefont{K.}~\bibnamefont{Ueki}},
  \bibinfo{author}{\bibfnamefont{A.}~\bibnamefont{Tokiwa}},
  \bibinfo{author}{\bibfnamefont{M.}~\bibnamefont{Kiruchi}},
  \bibinfo{author}{\bibfnamefont{Y.}~\bibnamefont{Syono}}, \bibnamefont{and}
  \bibinfo{author}{\bibfnamefont{M.}~\bibnamefont{Tachiki}},
  \bibinfo{journal}{Journal of Physics: Condensed Matter}
  \textbf{\bibinfo{volume}{4}}, \bibinfo{pages}{5769} (\bibinfo{year}{1992}).

\bibitem[{\citenamefont{Foyevtsova et~al.}(2015)\citenamefont{Foyevtsova,
  Khazraie, Elfimov, and Sawatzky}}]{foy15}
\bibinfo{author}{\bibfnamefont{K.}~\bibnamefont{Foyevtsova}},
  \bibinfo{author}{\bibfnamefont{A.}~\bibnamefont{Khazraie}},
  \bibinfo{author}{\bibfnamefont{I.}~\bibnamefont{Elfimov}}, \bibnamefont{and}
  \bibinfo{author}{\bibfnamefont{G.~A.} \bibnamefont{Sawatzky}},
  \bibinfo{journal}{Phys. Rev. B} \textbf{\bibinfo{volume}{91}},
  \bibinfo{pages}{121114} (\bibinfo{year}{2015}).

\bibitem[{\citenamefont{Plumb et~al.}(2016)\citenamefont{Plumb, Gawryluk, Wang,
  Risti\ifmmode~\acute{c}\else \'{c}\fi{}, Park, Lv, Wang, Matt, Xu, Shang
  et~al.}}]{plumb16}
\bibinfo{author}{\bibfnamefont{N.~C.} \bibnamefont{Plumb}},
  \bibinfo{author}{\bibfnamefont{D.~J.} \bibnamefont{Gawryluk}},
  \bibinfo{author}{\bibfnamefont{Y.}~\bibnamefont{Wang}},
  \bibinfo{author}{\bibfnamefont{Z.}~\bibnamefont{Risti\ifmmode~\acute{c}\else
  \'{c}\fi{}}}, \bibinfo{author}{\bibfnamefont{J.}~\bibnamefont{Park}},
  \bibinfo{author}{\bibfnamefont{B.~Q.} \bibnamefont{Lv}},
  \bibinfo{author}{\bibfnamefont{Z.}~\bibnamefont{Wang}},
  \bibinfo{author}{\bibfnamefont{C.~E.} \bibnamefont{Matt}},
  \bibinfo{author}{\bibfnamefont{N.}~\bibnamefont{Xu}},
  \bibinfo{author}{\bibfnamefont{T.}~\bibnamefont{Shang}},
  \bibnamefont{et~al.}, \bibinfo{journal}{Phys. Rev. Lett.}
  \textbf{\bibinfo{volume}{117}}, \bibinfo{pages}{037002}
  (\bibinfo{year}{2016}).

\bibitem[{\citenamefont{Balandeh et~al.}(2017)\citenamefont{Balandeh, Green,
  Foyevtsova, Chi, Foyevtsov, Li, and Sawatzky}}]{balan17}
\bibinfo{author}{\bibfnamefont{S.}~\bibnamefont{Balandeh}},
  \bibinfo{author}{\bibfnamefont{R.~J.} \bibnamefont{Green}},
  \bibinfo{author}{\bibfnamefont{K.}~\bibnamefont{Foyevtsova}},
  \bibinfo{author}{\bibfnamefont{S.}~\bibnamefont{Chi}},
  \bibinfo{author}{\bibfnamefont{O.}~\bibnamefont{Foyevtsov}},
  \bibinfo{author}{\bibfnamefont{F.}~\bibnamefont{Li}}, \bibnamefont{and}
  \bibinfo{author}{\bibfnamefont{G.~A.} \bibnamefont{Sawatzky}},
  \bibinfo{journal}{Phys. Rev. B} \textbf{\bibinfo{volume}{96}},
  \bibinfo{pages}{165127} (\bibinfo{year}{2017}).

\bibitem[{\citenamefont{Yan et~al.}(2013)\citenamefont{Yan, Jansen, and
  Felser}}]{yan13}
\bibinfo{author}{\bibfnamefont{B.}~\bibnamefont{Yan}},
  \bibinfo{author}{\bibfnamefont{M.}~\bibnamefont{Jansen}}, \bibnamefont{and}
  \bibinfo{author}{\bibfnamefont{C.}~\bibnamefont{Felser}},
  \bibinfo{journal}{Nature Phys.} \textbf{\bibinfo{volume}{9}},
  \bibinfo{pages}{709} (\bibinfo{year}{2013}).

\bibitem[{\citenamefont{Vildosola et~al.}(2013)\citenamefont{Vildosola,
  Güller, and Llois}}]{Vildosola2013}
\bibinfo{author}{\bibfnamefont{V.}~\bibnamefont{Vildosola}},
  \bibinfo{author}{\bibfnamefont{F.}~\bibnamefont{Güller}}, \bibnamefont{and}
  \bibinfo{author}{\bibfnamefont{A.~M.} \bibnamefont{Llois}},
  \bibinfo{journal}{Physical Review Letters} \textbf{\bibinfo{volume}{110}},
  \bibinfo{pages}{206805} (\bibinfo{year}{2013}).

\bibitem[{\citenamefont{Meir et~al.}(2017)\citenamefont{Meir, Gorol, Kopp, and
  Hammerl}}]{Meir2017}
\bibinfo{author}{\bibfnamefont{B.}~\bibnamefont{Meir}},
  \bibinfo{author}{\bibfnamefont{S.}~\bibnamefont{Gorol}},
  \bibinfo{author}{\bibfnamefont{T.}~\bibnamefont{Kopp}}, \bibnamefont{and}
  \bibinfo{author}{\bibfnamefont{G.}~\bibnamefont{Hammerl}},
  \bibinfo{journal}{Phys. Rev. B} \textbf{\bibinfo{volume}{96}},
  \bibinfo{pages}{R100507} (\bibinfo{year}{2017}).

\bibitem[{\citenamefont{Di~Napoli et~al.}(2021)\citenamefont{Di~Napoli, Helman,
  Llois, and Vildosola}}]{arxiv-BPO-BBO}
\bibinfo{author}{\bibfnamefont{S.}~\bibnamefont{Di~Napoli}},
  \bibinfo{author}{\bibfnamefont{C.}~\bibnamefont{Helman}},
  \bibinfo{author}{\bibfnamefont{A.~M.} \bibnamefont{Llois}}, \bibnamefont{and}
  \bibinfo{author}{\bibfnamefont{V.}~\bibnamefont{Vildosola}},
  \bibinfo{journal}{arxiv: 2011.08710}  (\bibinfo{year}{2021}).

\bibitem[{\citenamefont{Makita and Abe}(1997)}]{makita97}
\bibinfo{author}{\bibfnamefont{T.}~\bibnamefont{Makita}} \bibnamefont{and}
  \bibinfo{author}{\bibfnamefont{H.}~\bibnamefont{Abe}},
  \bibinfo{journal}{Japanese Journal of Applied Physics}
  \textbf{\bibinfo{volume}{36}}, \bibinfo{pages}{L96} (\bibinfo{year}{1997}).

\bibitem[{\citenamefont{Gozar et~al.}(2007)\citenamefont{Gozar, Logvenov,
  Butko, and Bozovic}}]{gozar07}
\bibinfo{author}{\bibfnamefont{A.}~\bibnamefont{Gozar}},
  \bibinfo{author}{\bibfnamefont{G.}~\bibnamefont{Logvenov}},
  \bibinfo{author}{\bibfnamefont{V.~Y.} \bibnamefont{Butko}}, \bibnamefont{and}
  \bibinfo{author}{\bibfnamefont{I.}~\bibnamefont{Bozovic}},
  \bibinfo{journal}{Phys. Rev. B} \textbf{\bibinfo{volume}{75}},
  \bibinfo{pages}{201402} (\bibinfo{year}{2007}).

\bibitem[{\citenamefont{Inumaru et~al.}(2008)\citenamefont{Inumaru, Miyata, and
  Yamanaka}}]{inumaru08}
\bibinfo{author}{\bibfnamefont{K.}~\bibnamefont{Inumaru}},
  \bibinfo{author}{\bibfnamefont{H.}~\bibnamefont{Miyata}}, \bibnamefont{and}
  \bibinfo{author}{\bibfnamefont{S.}~\bibnamefont{Yamanaka}},
  \bibinfo{journal}{Phys. Rev. B} \textbf{\bibinfo{volume}{78}},
  \bibinfo{pages}{132507} (\bibinfo{year}{2008}).

\bibitem[{\citenamefont{Lee et~al.}(2016)\citenamefont{Lee, Kim, Hwang, Kim,
  Kang, Kim, Kim, and Noh}}]{lee16}
\bibinfo{author}{\bibfnamefont{H.~G.} \bibnamefont{Lee}},
  \bibinfo{author}{\bibfnamefont{Y.}~\bibnamefont{Kim}},
  \bibinfo{author}{\bibfnamefont{S.}~\bibnamefont{Hwang}},
  \bibinfo{author}{\bibfnamefont{G.}~\bibnamefont{Kim}},
  \bibinfo{author}{\bibfnamefont{T.~D.} \bibnamefont{Kang}},
  \bibinfo{author}{\bibfnamefont{M.}~\bibnamefont{Kim}},
  \bibinfo{author}{\bibfnamefont{M.}~\bibnamefont{Kim}}, \bibnamefont{and}
  \bibinfo{author}{\bibfnamefont{T.~W.} \bibnamefont{Noh}},
  \bibinfo{journal}{APL Materials} \textbf{\bibinfo{volume}{4}},
  \bibinfo{pages}{126106} (\bibinfo{year}{2016}).

\bibitem[{\citenamefont{Ferreyra
  et~al.}(2016{\natexlab{a}})\citenamefont{Ferreyra, Guller, Marchini, Lüders,
  Albornoz, Leyva, Williams, Llois, Vildosola, and Rubi}}]{ferreyra16}
\bibinfo{author}{\bibfnamefont{C.}~\bibnamefont{Ferreyra}},
  \bibinfo{author}{\bibfnamefont{F.}~\bibnamefont{Guller}},
  \bibinfo{author}{\bibfnamefont{F.}~\bibnamefont{Marchini}},
  \bibinfo{author}{\bibfnamefont{U.}~\bibnamefont{Lüders}},
  \bibinfo{author}{\bibfnamefont{C.}~\bibnamefont{Albornoz}},
  \bibinfo{author}{\bibfnamefont{A.~G.} \bibnamefont{Leyva}},
  \bibinfo{author}{\bibfnamefont{F.~J.} \bibnamefont{Williams}},
  \bibinfo{author}{\bibfnamefont{A.~M.} \bibnamefont{Llois}},
  \bibinfo{author}{\bibfnamefont{V.}~\bibnamefont{Vildosola}},
  \bibnamefont{and} \bibinfo{author}{\bibfnamefont{D.}~\bibnamefont{Rubi}},
  \bibinfo{journal}{AIP Advances} \textbf{\bibinfo{volume}{6}},
  \bibinfo{pages}{065310} (\bibinfo{year}{2016}{\natexlab{a}}).

\bibitem[{\citenamefont{Ferreyra
  et~al.}(2016{\natexlab{b}})\citenamefont{Ferreyra, Marchini, Granell, Golmar,
  Albornoz, Williams, Leyva, and Rubi}}]{ferreyra16b}
\bibinfo{author}{\bibfnamefont{C.}~\bibnamefont{Ferreyra}},
  \bibinfo{author}{\bibfnamefont{F.}~\bibnamefont{Marchini}},
  \bibinfo{author}{\bibfnamefont{P.}~\bibnamefont{Granell}},
  \bibinfo{author}{\bibfnamefont{F.}~\bibnamefont{Golmar}},
  \bibinfo{author}{\bibfnamefont{C.}~\bibnamefont{Albornoz}},
  \bibinfo{author}{\bibfnamefont{F.}~\bibnamefont{Williams}},
  \bibinfo{author}{\bibfnamefont{A.}~\bibnamefont{Leyva}}, \bibnamefont{and}
  \bibinfo{author}{\bibfnamefont{D.}~\bibnamefont{Rubi}},
  \bibinfo{journal}{Thin Solid Films} \textbf{\bibinfo{volume}{612}},
  \bibinfo{pages}{369} (\bibinfo{year}{2016}{\natexlab{b}}).

\bibitem[{\citenamefont{Zapf et~al.}(2018)\citenamefont{Zapf, Stübinger, Jin,
  Kamp, Pfaff, Lubk, Büchner, Sing, and Claessen}}]{zapf18}
\bibinfo{author}{\bibfnamefont{M.}~\bibnamefont{Zapf}},
  \bibinfo{author}{\bibfnamefont{M.}~\bibnamefont{Stübinger}},
  \bibinfo{author}{\bibfnamefont{L.}~\bibnamefont{Jin}},
  \bibinfo{author}{\bibfnamefont{M.}~\bibnamefont{Kamp}},
  \bibinfo{author}{\bibfnamefont{F.}~\bibnamefont{Pfaff}},
  \bibinfo{author}{\bibfnamefont{A.}~\bibnamefont{Lubk}},
  \bibinfo{author}{\bibfnamefont{B.}~\bibnamefont{Büchner}},
  \bibinfo{author}{\bibfnamefont{M.}~\bibnamefont{Sing}}, \bibnamefont{and}
  \bibinfo{author}{\bibfnamefont{R.}~\bibnamefont{Claessen}},
  \bibinfo{journal}{Applied Physics Letters} \textbf{\bibinfo{volume}{112}},
  \bibinfo{pages}{141601} (\bibinfo{year}{2018}).

\bibitem[{\citenamefont{Zapf et~al.}(2019)\citenamefont{Zapf, Els\"asser,
  St\"ubinger, Scheiderer, Geurts, Sing, and Claessen}}]{zapf19}
\bibinfo{author}{\bibfnamefont{M.}~\bibnamefont{Zapf}},
  \bibinfo{author}{\bibfnamefont{S.}~\bibnamefont{Els\"asser}},
  \bibinfo{author}{\bibfnamefont{M.}~\bibnamefont{St\"ubinger}},
  \bibinfo{author}{\bibfnamefont{P.}~\bibnamefont{Scheiderer}},
  \bibinfo{author}{\bibfnamefont{J.}~\bibnamefont{Geurts}},
  \bibinfo{author}{\bibfnamefont{M.}~\bibnamefont{Sing}}, \bibnamefont{and}
  \bibinfo{author}{\bibfnamefont{R.}~\bibnamefont{Claessen}},
  \bibinfo{journal}{Phys. Rev. B} \textbf{\bibinfo{volume}{99}},
  \bibinfo{pages}{245308} (\bibinfo{year}{2019}).

\bibitem[{\citenamefont{Bouwmeester et~al.}(2019)\citenamefont{Bouwmeester,
  de~Hond, Gauquelin, Verbeeck, Koster, and Brinkman}}]{bouwmeester19}
\bibinfo{author}{\bibfnamefont{R.~L.} \bibnamefont{Bouwmeester}},
  \bibinfo{author}{\bibfnamefont{K.}~\bibnamefont{de~Hond}},
  \bibinfo{author}{\bibfnamefont{N.}~\bibnamefont{Gauquelin}},
  \bibinfo{author}{\bibfnamefont{J.}~\bibnamefont{Verbeeck}},
  \bibinfo{author}{\bibfnamefont{G.}~\bibnamefont{Koster}}, \bibnamefont{and}
  \bibinfo{author}{\bibfnamefont{A.}~\bibnamefont{Brinkman}},
  \bibinfo{journal}{physica status solidi (RRL)} \textbf{\bibinfo{volume}{13}},
  \bibinfo{pages}{1800679} (\bibinfo{year}{2019}).

\bibitem[{\citenamefont{Jin et~al.}(2020)\citenamefont{Jin, Zapf, Stübinger,
  Kamp, Sing, Claessen, and Jia}}]{jin2020}
\bibinfo{author}{\bibfnamefont{L.}~\bibnamefont{Jin}},
  \bibinfo{author}{\bibfnamefont{M.}~\bibnamefont{Zapf}},
  \bibinfo{author}{\bibfnamefont{M.}~\bibnamefont{Stübinger}},
  \bibinfo{author}{\bibfnamefont{M.}~\bibnamefont{Kamp}},
  \bibinfo{author}{\bibfnamefont{M.}~\bibnamefont{Sing}},
  \bibinfo{author}{\bibfnamefont{R.}~\bibnamefont{Claessen}}, \bibnamefont{and}
  \bibinfo{author}{\bibfnamefont{C.-L.} \bibnamefont{Jia}},
  \bibinfo{journal}{physica status solidi (RRL)} \textbf{\bibinfo{volume}{14}},
  \bibinfo{pages}{2000054} (\bibinfo{year}{2020}).

\bibitem[{\citenamefont{Kim et~al.}(2015)\citenamefont{Kim, Neumann, Kim, Le,
  Kang, and Noh}}]{kim15}
\bibinfo{author}{\bibfnamefont{G.}~\bibnamefont{Kim}},
  \bibinfo{author}{\bibfnamefont{M.}~\bibnamefont{Neumann}},
  \bibinfo{author}{\bibfnamefont{M.}~\bibnamefont{Kim}},
  \bibinfo{author}{\bibfnamefont{M.~D.} \bibnamefont{Le}},
  \bibinfo{author}{\bibfnamefont{T.~D.} \bibnamefont{Kang}}, \bibnamefont{and}
  \bibinfo{author}{\bibfnamefont{T.~W.} \bibnamefont{Noh}},
  \bibinfo{journal}{Phys. Rev. Lett.} \textbf{\bibinfo{volume}{115}},
  \bibinfo{pages}{226402} (\bibinfo{year}{2015}).

\bibitem[{\citenamefont{Skinner and Kilner}(2003)}]{skinner2003}
\bibinfo{author}{\bibfnamefont{S.~J.} \bibnamefont{Skinner}} \bibnamefont{and}
  \bibinfo{author}{\bibfnamefont{J.~A.} \bibnamefont{Kilner}},
  \bibinfo{journal}{Materials Today} \textbf{\bibinfo{volume}{6}},
  \bibinfo{pages}{30} (\bibinfo{year}{2003}).

\bibitem[{\citenamefont{Kilo et~al.}(2000)\citenamefont{Kilo, Borchardt,
  Lesage, Ka\"itasov, Weber, and Scherrer}}]{kilo2000}
\bibinfo{author}{\bibfnamefont{M.}~\bibnamefont{Kilo}},
  \bibinfo{author}{\bibfnamefont{G.}~\bibnamefont{Borchardt}},
  \bibinfo{author}{\bibfnamefont{B.}~\bibnamefont{Lesage}},
  \bibinfo{author}{\bibfnamefont{O.}~\bibnamefont{Ka\"itasov}},
  \bibinfo{author}{\bibfnamefont{S.}~\bibnamefont{Weber}}, \bibnamefont{and}
  \bibinfo{author}{\bibfnamefont{S.}~\bibnamefont{Scherrer}},
  \bibinfo{journal}{Journal of the European Ceramic Society}
  \textbf{\bibinfo{volume}{20}}, \bibinfo{pages}{2069} (\bibinfo{year}{2000}).

\bibitem[{SI()}]{SI}
\bibinfo{journal}{See Supplemental Material at (link) for additional X-ray
  diffraction and TEM experiments}  (????).

\bibitem[{\citenamefont{Keeble et~al.}(2013)\citenamefont{Keeble, Wicklein,
  Jin, Jia, Egger, and Dittmann}}]{keeble13}
\bibinfo{author}{\bibfnamefont{D.~J.} \bibnamefont{Keeble}},
  \bibinfo{author}{\bibfnamefont{S.}~\bibnamefont{Wicklein}},
  \bibinfo{author}{\bibfnamefont{L.}~\bibnamefont{Jin}},
  \bibinfo{author}{\bibfnamefont{C.~L.} \bibnamefont{Jia}},
  \bibinfo{author}{\bibfnamefont{W.}~\bibnamefont{Egger}}, \bibnamefont{and}
  \bibinfo{author}{\bibfnamefont{R.}~\bibnamefont{Dittmann}},
  \bibinfo{journal}{Phys. Rev. B} \textbf{\bibinfo{volume}{87}},
  \bibinfo{pages}{195409} (\bibinfo{year}{2013}).

\bibitem[{\citenamefont{Brooks et~al.}(2015)\citenamefont{Brooks, Wilson,
  Schäfer, Mundy, Holtz, Muller, Schubert, Cahill, and Schlom}}]{brooks15}
\bibinfo{author}{\bibfnamefont{C.~M.} \bibnamefont{Brooks}},
  \bibinfo{author}{\bibfnamefont{R.~B.} \bibnamefont{Wilson}},
  \bibinfo{author}{\bibfnamefont{A.}~\bibnamefont{Schäfer}},
  \bibinfo{author}{\bibfnamefont{J.~A.} \bibnamefont{Mundy}},
  \bibinfo{author}{\bibfnamefont{M.~E.} \bibnamefont{Holtz}},
  \bibinfo{author}{\bibfnamefont{D.~A.} \bibnamefont{Muller}},
  \bibinfo{author}{\bibfnamefont{J.}~\bibnamefont{Schubert}},
  \bibinfo{author}{\bibfnamefont{D.~G.} \bibnamefont{Cahill}},
  \bibnamefont{and} \bibinfo{author}{\bibfnamefont{D.~G.}
  \bibnamefont{Schlom}}, \bibinfo{journal}{Applied Physics Letters}
  \textbf{\bibinfo{volume}{107}}, \bibinfo{pages}{051902}
  (\bibinfo{year}{2015}).

\bibitem[{\citenamefont{Scafetta and May}(2017)}]{scafetta17}
\bibinfo{author}{\bibfnamefont{M.~D.} \bibnamefont{Scafetta}} \bibnamefont{and}
  \bibinfo{author}{\bibfnamefont{S.~J.} \bibnamefont{May}},
  \bibinfo{journal}{Phys. Chem. Chem. Phys.} \textbf{\bibinfo{volume}{19}},
  \bibinfo{pages}{10371} (\bibinfo{year}{2017}).

\bibitem[{\citenamefont{Li et~al.}(2019)\citenamefont{Li, Davidson, Sutarto,
  Shin, Liu, Elfimov, Foyevtsova, He, Sawatzky, and Zou}}]{li2019}
\bibinfo{author}{\bibfnamefont{F.}~\bibnamefont{Li}},
  \bibinfo{author}{\bibfnamefont{B.~A.} \bibnamefont{Davidson}},
  \bibinfo{author}{\bibfnamefont{R.}~\bibnamefont{Sutarto}},
  \bibinfo{author}{\bibfnamefont{H.}~\bibnamefont{Shin}},
  \bibinfo{author}{\bibfnamefont{C.}~\bibnamefont{Liu}},
  \bibinfo{author}{\bibfnamefont{I.}~\bibnamefont{Elfimov}},
  \bibinfo{author}{\bibfnamefont{K.}~\bibnamefont{Foyevtsova}},
  \bibinfo{author}{\bibfnamefont{F.}~\bibnamefont{He}},
  \bibinfo{author}{\bibfnamefont{G.~A.} \bibnamefont{Sawatzky}},
  \bibnamefont{and} \bibinfo{author}{\bibfnamefont{K.}~\bibnamefont{Zou}},
  \bibinfo{journal}{Phys. Rev. Materials} \textbf{\bibinfo{volume}{3}},
  \bibinfo{pages}{100802} (\bibinfo{year}{2019}).

\bibitem[{\citenamefont{Sleight}(2015)}]{sleight15}
\bibinfo{author}{\bibfnamefont{A.~W.} \bibnamefont{Sleight}},
  \bibinfo{journal}{Phys. C} \textbf{\bibinfo{volume}{514}},
  \bibinfo{pages}{152} (\bibinfo{year}{2015}).

\bibitem[{\citenamefont{Prakash et~al.}(2016)\citenamefont{Prakash, Xu,
  Faghaninia, Shukla, Ager, Lo, and Jalan}}]{prakash16}
\bibinfo{author}{\bibfnamefont{A.}~\bibnamefont{Prakash}},
  \bibinfo{author}{\bibfnamefont{P.}~\bibnamefont{Xu}},
  \bibinfo{author}{\bibfnamefont{A.}~\bibnamefont{Faghaninia}},
  \bibinfo{author}{\bibfnamefont{S.}~\bibnamefont{Shukla}},
  \bibinfo{author}{\bibfnamefont{J.~W.} \bibnamefont{Ager}},
  \bibinfo{author}{\bibfnamefont{C.~S.} \bibnamefont{Lo}}, \bibnamefont{and}
  \bibinfo{author}{\bibfnamefont{B.}~\bibnamefont{Jalan}},
  \bibinfo{journal}{Nature Commun.} \textbf{\bibinfo{volume}{8}},
  \bibinfo{pages}{15167} (\bibinfo{year}{2016}).

\bibitem[{\citenamefont{Zhao et~al.}(2016)\citenamefont{Zhao, Shi, Xi, Wang,
  and Shuai}}]{zhao16}
\bibinfo{author}{\bibfnamefont{T.}~\bibnamefont{Zhao}},
  \bibinfo{author}{\bibfnamefont{W.}~\bibnamefont{Shi}},
  \bibinfo{author}{\bibfnamefont{J.}~\bibnamefont{Xi}},
  \bibinfo{author}{\bibfnamefont{D.}~\bibnamefont{Wang}}, \bibnamefont{and}
  \bibinfo{author}{\bibfnamefont{Z.}~\bibnamefont{Shuai}},
  \bibinfo{journal}{Sci. Reports} \textbf{\bibinfo{volume}{6}},
  \bibinfo{pages}{19968} (\bibinfo{year}{2016}).

\bibitem[{\citenamefont{Dolloff}(1956)}]{dollof56}
\bibinfo{author}{\bibfnamefont{R.~T.} \bibnamefont{Dolloff}},
  \bibinfo{journal}{Journal of Applied Physics} \textbf{\bibinfo{volume}{27}},
  \bibinfo{pages}{1418} (\bibinfo{year}{1956}).

\bibitem[{\citenamefont{Jaime et~al.}(1997)\citenamefont{Jaime, Hardner,
  Salamon, Rubinstein, Dorsey, and Emin}}]{jaime97}
\bibinfo{author}{\bibfnamefont{M.}~\bibnamefont{Jaime}},
  \bibinfo{author}{\bibfnamefont{H.~T.} \bibnamefont{Hardner}},
  \bibinfo{author}{\bibfnamefont{M.~B.} \bibnamefont{Salamon}},
  \bibinfo{author}{\bibfnamefont{M.}~\bibnamefont{Rubinstein}},
  \bibinfo{author}{\bibfnamefont{P.}~\bibnamefont{Dorsey}}, \bibnamefont{and}
  \bibinfo{author}{\bibfnamefont{D.}~\bibnamefont{Emin}},
  \bibinfo{journal}{Phys. Rev. Lett.} \textbf{\bibinfo{volume}{78}},
  \bibinfo{pages}{951} (\bibinfo{year}{1997}).

\bibitem[{\citenamefont{Sugai et~al.}(1985)\citenamefont{Sugai, Uchida,
  Kitazawa, Tanaka, and Katsui}}]{PhysRevLett.55.426}
\bibinfo{author}{\bibfnamefont{S.}~\bibnamefont{Sugai}},
  \bibinfo{author}{\bibfnamefont{S.}~\bibnamefont{Uchida}},
  \bibinfo{author}{\bibfnamefont{K.}~\bibnamefont{Kitazawa}},
  \bibinfo{author}{\bibfnamefont{S.}~\bibnamefont{Tanaka}}, \bibnamefont{and}
  \bibinfo{author}{\bibfnamefont{A.}~\bibnamefont{Katsui}},
  \bibinfo{journal}{Phys. Rev. Lett.} \textbf{\bibinfo{volume}{55}},
  \bibinfo{pages}{426} (\bibinfo{year}{1985}).

\bibitem[{\citenamefont{Franchini et~al.}(2009)\citenamefont{Franchini, Kresse,
  and Podloucky}}]{franchini09}
\bibinfo{author}{\bibfnamefont{C.}~\bibnamefont{Franchini}},
  \bibinfo{author}{\bibfnamefont{G.}~\bibnamefont{Kresse}}, \bibnamefont{and}
  \bibinfo{author}{\bibfnamefont{R.}~\bibnamefont{Podloucky}},
  \bibinfo{journal}{Phys. Rev. Lett.} \textbf{\bibinfo{volume}{102}},
  \bibinfo{pages}{256402} (\bibinfo{year}{2009}).

\bibitem[{\citenamefont{Bl\"ochl}(1994)}]{PAW}
\bibinfo{author}{\bibfnamefont{P.}~\bibnamefont{Bl\"ochl}},
  \bibinfo{journal}{Phys. Rev. B} \textbf{\bibinfo{volume}{50}},
  \bibinfo{pages}{17953} (\bibinfo{year}{1994}).

\bibitem[{\citenamefont{Kresse and Furthm\"uller}(1996)}]{VASP}
\bibinfo{author}{\bibfnamefont{G.}~\bibnamefont{Kresse}} \bibnamefont{and}
  \bibinfo{author}{\bibfnamefont{J.}~\bibnamefont{Furthm\"uller}},
  \bibinfo{journal}{Phys. Rev. B} \textbf{\bibinfo{volume}{54}},
  \bibinfo{pages}{11169} (\bibinfo{year}{1996}).

\bibitem[{\citenamefont{Kresse and Joubert}(1999)}]{PAW-VASP}
\bibinfo{author}{\bibfnamefont{G.}~\bibnamefont{Kresse}} \bibnamefont{and}
  \bibinfo{author}{\bibfnamefont{D.}~\bibnamefont{Joubert}},
  \bibinfo{journal}{Phys. Rev. B} \textbf{\bibinfo{volume}{59}},
  \bibinfo{pages}{1758} (\bibinfo{year}{1999}).

\bibitem[{\citenamefont{Perdew et~al.}(1996)\citenamefont{Perdew, Burke, and
  M}}]{PBE96}
\bibinfo{author}{\bibfnamefont{J.}~\bibnamefont{Perdew}},
  \bibinfo{author}{\bibfnamefont{S.}~\bibnamefont{Burke}}, \bibnamefont{and}
  \bibinfo{author}{\bibfnamefont{M.~E.} \bibnamefont{M}},
  \bibinfo{journal}{Phys. Rev. Lett.} \textbf{\bibinfo{volume}{77}},
  \bibinfo{pages}{3865} (\bibinfo{year}{1996}).

\bibitem[{\citenamefont{Heyd et~al.}(2003)\citenamefont{Heyd, Scuseria, and
  Ernzerhof}}]{2003-HSE1}
\bibinfo{author}{\bibfnamefont{J.}~\bibnamefont{Heyd}},
  \bibinfo{author}{\bibfnamefont{G.~E.} \bibnamefont{Scuseria}},
  \bibnamefont{and}
  \bibinfo{author}{\bibfnamefont{M.}~\bibnamefont{Ernzerhof}},
  \bibinfo{journal}{The Journal of Chemical Physics}
  \textbf{\bibinfo{volume}{118}}, \bibinfo{pages}{8207} (\bibinfo{year}{2003}).

\bibitem[{\citenamefont{Krukau et~al.}(2006)\citenamefont{Krukau, Vydrov,
  Izmaylov, and Scuseria}}]{2006-HSE2}
\bibinfo{author}{\bibfnamefont{A.~V.} \bibnamefont{Krukau}},
  \bibinfo{author}{\bibfnamefont{O.~A.} \bibnamefont{Vydrov}},
  \bibinfo{author}{\bibfnamefont{A.~F.} \bibnamefont{Izmaylov}},
  \bibnamefont{and} \bibinfo{author}{\bibfnamefont{G.~E.}
  \bibnamefont{Scuseria}}, \bibinfo{journal}{The Journal of Chemical Physics}
  \textbf{\bibinfo{volume}{125}}, \bibinfo{pages}{224106}
  (\bibinfo{year}{2006}).

\bibitem[{\citenamefont{Cox and Sleight}(1976{\natexlab{b}})}]{Cox1976969}
\bibinfo{author}{\bibfnamefont{D.}~\bibnamefont{Cox}} \bibnamefont{and}
  \bibinfo{author}{\bibfnamefont{A.}~\bibnamefont{Sleight}},
  \bibinfo{journal}{Solid State Communications} \textbf{\bibinfo{volume}{19}},
  \bibinfo{pages}{969 } (\bibinfo{year}{1976}{\natexlab{b}}).

\bibitem[{\citenamefont{Franchini et~al.}(2010)\citenamefont{Franchini, Sanna,
  Marsman, and Kresse}}]{PhysRevB.81.085213}
\bibinfo{author}{\bibfnamefont{C.}~\bibnamefont{Franchini}},
  \bibinfo{author}{\bibfnamefont{A.}~\bibnamefont{Sanna}},
  \bibinfo{author}{\bibfnamefont{M.}~\bibnamefont{Marsman}}, \bibnamefont{and}
  \bibinfo{author}{\bibfnamefont{G.}~\bibnamefont{Kresse}},
  \bibinfo{journal}{Phys. Rev. B} \textbf{\bibinfo{volume}{81}},
  \bibinfo{pages}{085213} (\bibinfo{year}{2010}).

\bibitem[{\citenamefont{Henkelman and Jónsson}(2000)}]{NEB1}
\bibinfo{author}{\bibfnamefont{G.}~\bibnamefont{Henkelman}} \bibnamefont{and}
  \bibinfo{author}{\bibfnamefont{H.}~\bibnamefont{Jónsson}},
  \bibinfo{journal}{J. Chem. Phys.} \textbf{\bibinfo{volume}{113}},
  \bibinfo{pages}{9901} (\bibinfo{year}{2000}).

\bibitem[{\citenamefont{Yin et~al.}(2013)\citenamefont{Yin, Kutepov, and
  Kotliar}}]{Kotliar2013}
\bibinfo{author}{\bibfnamefont{Z.~P.} \bibnamefont{Yin}},
  \bibinfo{author}{\bibfnamefont{A.}~\bibnamefont{Kutepov}}, \bibnamefont{and}
  \bibinfo{author}{\bibfnamefont{G.}~\bibnamefont{Kotliar}},
  \bibinfo{journal}{Physical Review X} \textbf{\bibinfo{volume}{3}}
  (\bibinfo{year}{2013}).

\bibitem[{\citenamefont{Tilley}(2008)}]{defects-in-solids}
\bibinfo{author}{\bibfnamefont{R.~J.~D.} \bibnamefont{Tilley}},
  \emph{\bibinfo{title}{Defects in solids}} (\bibinfo{publisher}{John Wiley \&
  Sons, Inc}, \bibinfo{year}{2008}).

\end{thebibliography}


\begin{thebibliography}{0}
\expandafter\ifx\csname natexlab\endcsname\relax\def\natexlab#1{#1}\fi
\expandafter\ifx\csname bibnamefont\endcsname\relax
  \def\bibnamefont#1{#1}\fi
\expandafter\ifx\csname bibfnamefont\endcsname\relax
  \def\bibfnamefont#1{#1}\fi
\expandafter\ifx\csname citenamefont\endcsname\relax
  \def\citenamefont#1{#1}\fi
\expandafter\ifx\csname url\endcsname\relax
  \def\url#1{\texttt{#1}}\fi
\expandafter\ifx\csname urlprefix\endcsname\relax\def\urlprefix{URL }\fi
\providecommand{\bibinfo}[2]{#2}
\providecommand{\eprint}[2][]{\url{#2}}

\end{thebibliography}
\end{document}


\title{Supplemental Information: Polarons formation in Bi-deficient BaBiO$_3$}

\author{W. Rom\'an Acevedo}

\author{S. Di Napoli}

\author{F. Romano}

\author{G. Rodr\'{\i}guez Ruiz}

\author{P. Nukala}

\author{C. Quinteros}

\author{J. Lecourt}

\author{U. Lüders}

\author{V. Vildosola}

\author{D. Rubi}

\begin{abstract}
\end{abstract}

\maketitle

\section{Complementary X-ray characterization and analysis}

Figure S1 displays an X-ray spectrum recorded for the bare sample holder used to characterized the capped-BBO sample. It is found that the peaks marked with * in Figure 1(a) of the manuscript correspond to the sample holder. The bare BBO sample X-ray spectrum was recorded with a different sample holder which does not present these peaks.\\

Figure S2 displays fittings performed on bare and capped BBO (200)$_{PC}$ peaks, respectively. Bare BBO peak was fitted with a single Gaussian centered at 41.59º (Figure S2(a)), while the fitting of capped-BBO peak required two Gaussians, centered at 41.59 º and 41.25 º (Figure S2(b)). These two components correspond to BBO and Bi-deficient BBO layers, respectively.\\

\section{2.	Complementary TEM characterization }

Figure S3(left) displays EDS maps recorded for the capped-BBO sampled. Figure S3(right) shows EDS line-scans, evidencing a step down of the Bi intensity in the Bi-deficient layer. Figure S4 displays a HAADF image taken after the EDS measurements, evidencing some grain growth together with a loss of HAADF contrast between BBO and Bi-deficient BBO layers. This suggests Bi back-diffusion from YSZ to BBO upon EDS mapping. Also, some Ba and Zr intermixing between BBO and YSZ is observed in the exposed sample. With the mentioned constraint in mind, a Bi vacancies amount of at least 8-10 \% can be estimated for the Bi-deficient layer. 

\begin{figure*}
\renewcommand{\figurename}{Figure S}
\centering
\includegraphics[scale=1]{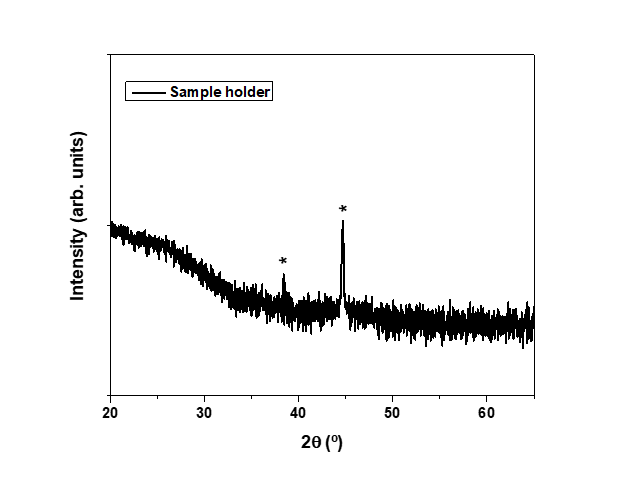}
\caption{X-ray diffraction spectrum recorded for the sample holder used to characterized the BBO-capped sample.}
\label{FigS1}
\end{figure*}

\begin{figure*}
\renewcommand{\figurename}{Figure S}
\centering
\includegraphics[scale=0.95]{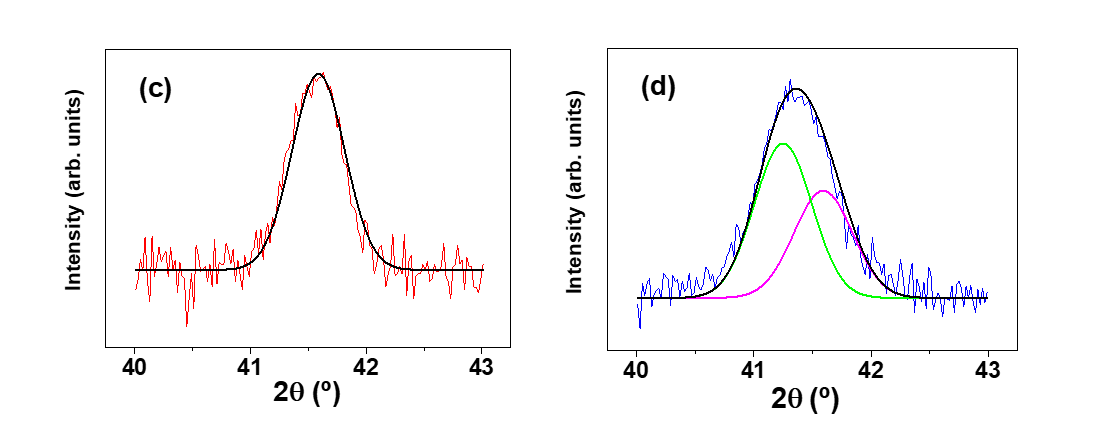}
\caption{X-ray (200)$_{PC}$ peaks –and the corresponding fittings- of (a) bare BBO and (b) capped-BBO samples.}
\label{FigS2}
\end{figure*}

\begin{figure*}
\renewcommand{\figurename}{Figure S}
\centering
\includegraphics[scale=0.65]{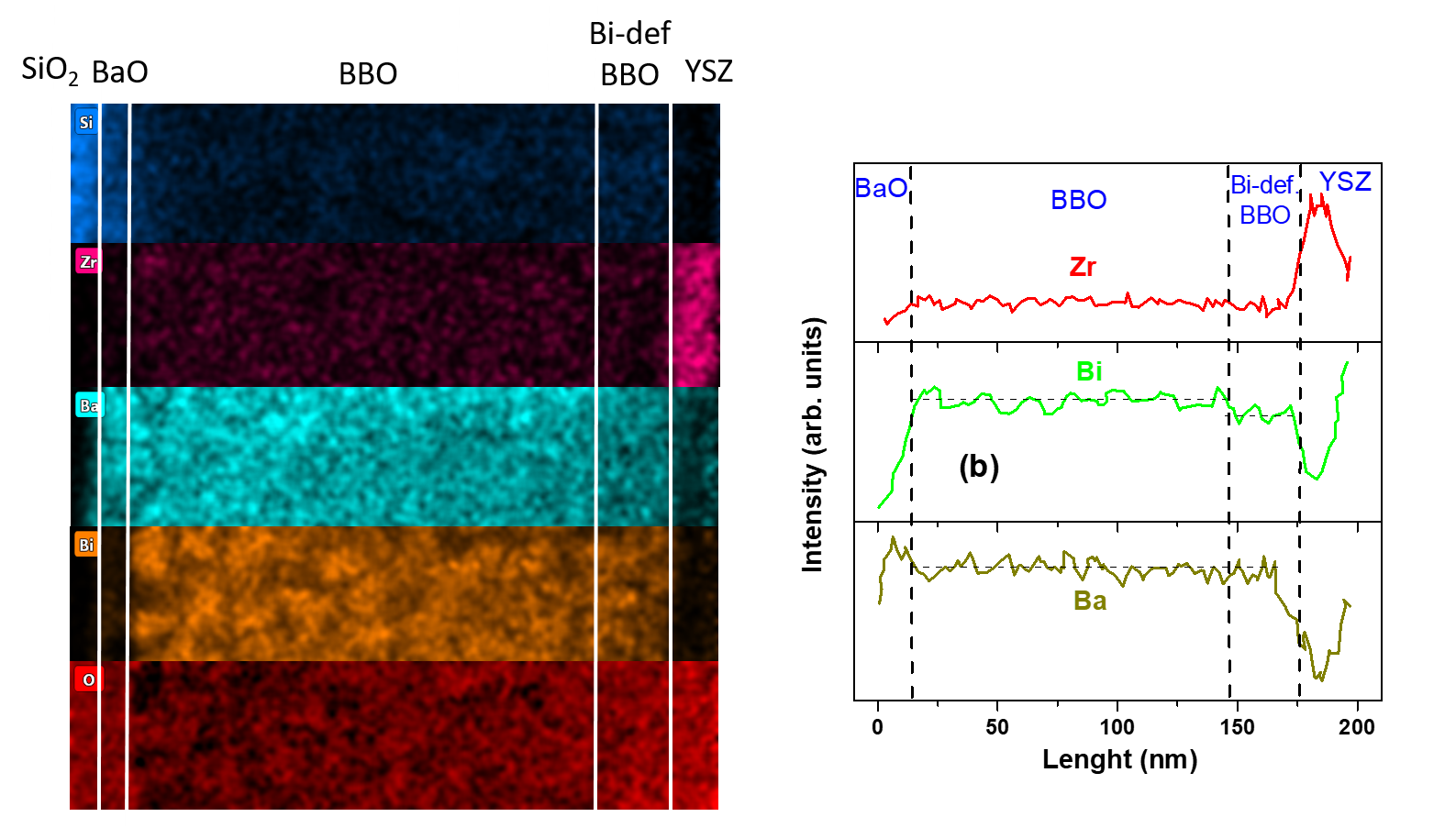}
\caption{(left) EDS maps for Si, Zr, Ba, Bi and O species performed in the capped-BBO sample; (right) EDS line-scans (intensity as a function of the position) for Zr, Bi and Ba for the same sample.}
\label{FigS3}
\end{figure*}

\begin{figure*}
\renewcommand{\figurename}{Figure S}
\centering
\includegraphics[scale=0.65]{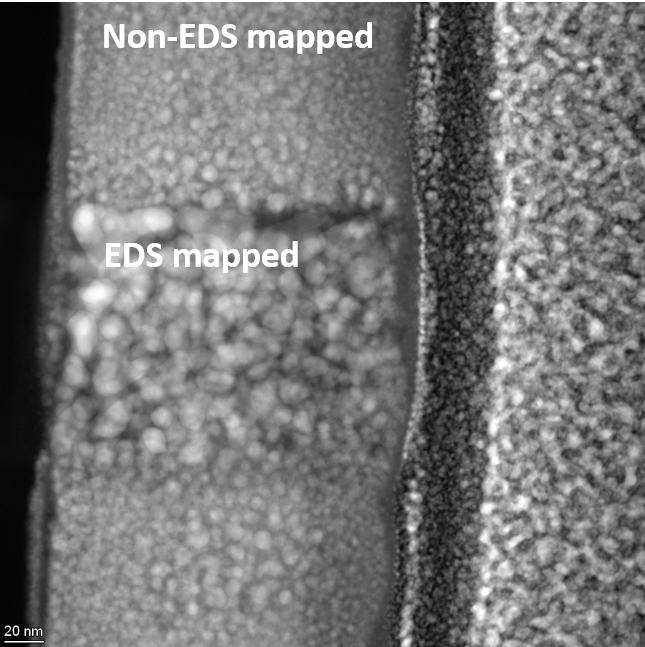}
\caption{HAADF image taken after exposing the capped-BBO sample to EDS mapping. Changes in the grain size and HAADF contrast are observed in the exposed region.}
\label{FigS4}
\end{figure*}

\bibliography{biblio_gases2D-oxidos}